\documentclass[useAMS,usenatbib]{mn2e}
\usepackage{subfigure}
\usepackage{natbib}
\usepackage{graphicx}
\usepackage{lscape}
\usepackage{rotating}
\usepackage{times}
\usepackage{epsf}
\usepackage{fancyhdr}
\usepackage{longtable,lscape}
\usepackage{amssymb,amsmath}

\newcommand{\xmmn}{{\it XMM-Newton~\/}}

\newcommand{\suzaku}{{\it Suzaku~\/}}

\newcommand{\ergscm}{erg\,cm$^{-2}$\,s$^{-1}$}

\newcommand{\ergs}{erg s$^{-1}$}

\def\eg{{e.g.~\/}}

\def\ie{{i.e.~\/}}

\title[The Galactic plane at faint X-ray fluxes]{The Galactic plane at faint X-ray fluxes - II. Stacked X-ray spectra of a sample of serendipitous {\it XMM-Newton} sources}
\author[R.~S. Warwick, K. Byckling, D. P\'erez-Ram\'irez]{R.~S. Warwick$^{1}$\thanks{E-mail:rsw@le.ac.uk}, K. Byckling$^{1}$, D. P\'erez-Ram\'irez$^{2}$ \\ 
$^{1}$Department of Physics and Astronomy, University of Leicester, University Road, Leicester, LE1 7RH, UK\\
$^{2}$Departamento de F\'isica, Universidad de Ja\'en, Campus Las Lagunillas, 23071 Ja\'en, Spain}

\begin{document}

\date{Accepted. Received; in original form}

\pagerange{\pageref{firstpage}--\pageref{lastpage}} \pubyear{2013}

\maketitle

\label{firstpage}

\begin{abstract}
We have investigated the X-ray spectral properties of a sample of 138 X-ray
sources detected serendipitously in \xmmn observations of the Galactic plane,
at an intermediate to faint flux level.  We divide our sample into 5
subgroups according to the spectral hardness of the sources, and stack 
(\ie co-add) the individual source spectra within each subgroup. 
As expected these stacked spectra show a softening trend from the hardest
to the softest subgroups, which is reflected in the inferred line-of-sight
column density. The spectra of the three hardest subgroups are characterized by
a hard continuum plus superimpose Fe-line emission in the 6--7~keV bandpass. 
The average equivalent width (EW) of the 6.7-keV He-like Fe-K$\alpha$ line 
is 170$^{+35}_{-32}$ eV, whereas the 6.4-keV Fe-K fluorescence line from neutral
iron and the 6.9-keV H-like Fe-Ly$\alpha$ line have EWs of 89$^{+26}_{-25}$ eV 
and 81$^{+30}_{-29}$ eV respectively, \ie roughly half that of the 6.7-keV line. 
The remaining subgroups exhibit soft thermal spectra.
Virtually all of the spectrally-soft X-ray sources can be associated with
relatively nearby coronally-active late-type stars, which are
evident as  bright near-infrared (NIR) objects within the X-ray error circles. 
On a similar basis only a minority of the spectrally-hard X-ray sources have
likely NIR identifications. The average continuum and Fe-line properties of the
spectrally-hard sources are consistent with those of magnetic
cataclysmic variables but the direct identification of large
numbers of such systems in Galactic X-ray surveys, probing intermediate to
faint flux levels, remains challenging. 

\end{abstract}


\section{Introduction}

From the ground breaking discoveries of missions, such as {\it Uhuru},
{\it SAS-3}, {\it Ariel V} and {\it HEAO-1}, that date
back over three decades, right through to the present era of observatory
class missions, such as {\it Chandra}, {\it XMM-Newton} and {\it Suzaku},
the study of X-ray bright Galactic sources has been a key theme within
high energy astrophysics.  As a result we now have extensive
knowledge of the most luminous X-ray emitters in our Galaxy 
(with X-ray luminosity typically in the range $10^{35-38}$ \ergs),
which in the main can be divided into either low-mass or high-mass
X-ray binaries (LMXBs or HMXBs) powered by the accretion of matter
onto either a neutron star or black hole (\eg \citealt{grimm02};
\citealt{ebi03}). At intermediate X-ray luminosities (nominally 
$10^{32-35}$ \ergs) other types of source in addition to X-ray binaries
enter the mix, including bright supernova remnants, cataclysmic
variables (CVs) powered by accretion onto a white dwarf star, massive stars 
with shock-heated winds and the most extreme coronally-active binaries, 
such as RS CVn systems (\eg \citealt{hertz84}; \citealt{motch10}).
Finally at relative low X-ray luminosities (L$_X$ $< 10^{32}$ \ergs)
the local population of coronally-active stars and binaries dominate
the source statistics, particularly in the soft X-ray regime 
(\eg \citealt{gudel04}).

Despite the great advances in our understanding of the various classes
of Galactic X-ray source, our knowledge of the population properties, 
such as the number density, the exact form of the luminosity function
and the Galactic distribution, remains very incomplete. This is of
particular relevance for modern day Galactic surveys conducted in the
hard X-ray band (\ie above 2 keV) where, for sources of intermediate
luminosity,  the visible volume can extend beyond the Galactic Centre
to encompass the bulk of the Galactic disc.  Ideally one would use
such surveys to construct large samples of identified sources from
which the properties of the various  populations might be inferred.
In  practice the identification of the counterparts to hard X-ray
sources in the Galactic plane, even with sub-arcsec X-ray positions,
is extraordinarily difficult due to the crowded nature of the star
fields and the effects of extinction at optical and near-infrared (NIR)
wavelengths (\eg \citealt{ebi05}; \citealt{laycock05}; \citealt{berg09}).  
Even when likely counterparts are identified, their follow-up, 
including distance determination, can often present significant
challenges (\eg \citealt{motch10}; \citealt{berg12}; \citealt{nebot13}).

One way of constraining the properties of low to intermediate
luminosity, source populations is through their integrated
emission. In the Galactic context, the spectrally hard
Galactic Ridge X-ray Emission (GRXE) (\citealt{wor82}; \citealt{war85};
\citealt{koy86a}; \citealt{yam96}) has been interpreted both in terms
of the superposition of faint point sources (\citealt{sug01};
\citealt{rev06}; \citealt{yuasa12}) and as a highly energetic, 
very high temperature phase of the interstellar medium 
(\citealt{koy86b}; \citealt{kan97}; \citealt{tanaka02}),
amongst other possibilities (\eg \citealt{val00}). The GRXE is seen as a
narrow ridge of emission
extending out to $ |l| \sim 60^{\circ}$, but with a surface
brightness that peaks towards the Galactic Centre (\citealt{yam93}; 
\citealt{rev06}; \citealt{koy07}). In the 4--10 keV band
the spectrum of the GRXE matches that of 5--10 keV optically-thin
CIE thermal plasma with prominent Fe-lines at 6.67 and 6.97 keV 
arising from K-shell emission in He-like and H-like ions 
(\citealt{koy86a}; \citealt{kan97}). An Fe K$\alpha$ line at 6.4 keV 
resulting from the fluorescence of neutral or near-neutral iron
is also evident in the X-ray spectrum of the GRXE
(\citealt{koy96}; \citealt{ebi08}; \citealt{yam09}). Hereafter we refer to the
three prominent Fe K$\alpha$  lines as the 6.4-keV, 6.7-keV and 6.9-keV
lines. Below 4 keV, emissions lines of abundant elements
such as Mg, Si, S, Ar and Ca are also evident in the spectrum of the GRXE,
with the intensity ratios of the K-lines associated with the
He-like and H-like ions of each element (including those of Fe)
indicative of a multi-temperature plasma (\citealt{kan97}; 
\citealt{tanaka02}). The temperature structure appears
to be similar at different locations along the GRXE, with the exception
of the region within a degree or so of the Galactic Centre
\citep{uch11,uch13}.

It has also been reported that above 10 keV the GRXE spectrum
exhibits a hard powerlaw tail, which extends to the hard 
X-ray/$\gamma$-ray region (\citealt{yam97}; \citealt{val98};
\citealt{strong05}; \citealt{kriv07}).

Recent studies have shown that the GRXE surface brightness follows
that of the NIR light associated with the old stellar population of the
Galaxy (\citealt{rev06}).  Also very deep {\it Chandra} observations
have directly resolved over 80\% of the GRXE near the Galactic
Centre into point
sources (\citealt{rev09}). Taken together this is compelling evidence
for the origin of the bulk of GRXE in the integrated emission of point
sources, although there is still some debate as to whether there might
remain some excess emission attributable to a distinct very hot diffuse
component within 2 degrees of the Galactic Centre (\citealt{uch11, uch13};
\citealt{heard13}; \citealt{nish13}).

The next key step
is to identify the Galactic source population or populations that give
rise to the GRXE. The first requirement, in this regard, is that the
source population should  have a sufficiently high  volume emissivity
(essentially the product of the mean space density and mean X-ray
luminosity) to explain the observed surface brightness of the GRXE.
A second constraint is that the integrated spectrum of the sources
should match the observed spectrum of the GRXE.
In this context it has been proposed that a mix of magnetic CVs plus 
coronally-active binaries may have sufficient spatial density and
hard X-ray luminosity to account for the bulk of the GRXE
and its extension into the Galactic Centre
(\citealt{muno04}; \citealt{saz06}; \citealt{rev06}; \citealt{rev08}). 
Also, CVs and active binaries have marked spectral similarities
to the GRXE (\citealt{rev06}; \citealt{tanaka10}; \citealt{yuasa12}).
However, as noted earlier, the current uncertainties
relating to the population properties and statistics leave
many of the details of this model to be confirmed.

In Warwick, P\'erez-Ram\'irez \& Byckling (2011; hereafter Paper I), 
we investigated the
serendipitous X-ray source population in the Galactic plane, 
utilising the Second {\it XMM-Newton} Serendipitous Source Catalogue
(2XMMi; \citealt{wat09}). The current paper
(Paper II) builds on the work in Paper I by focusing on
the X-ray spectral properties of a subset of sources
with relatively good photon statistics. In Paper I, 
we showed that the great majority
(\ie $>$90\%) of the spectrally-soft X-ray sources may be
identified with relatively local
coronally-active stars; however, the nature of the spectrally-hard
X-ray sources remains unclear.
The goal of the current paper is therefore to investigate the
X-ray spectral properties of a sample of sources representative
of the source population present in the Galactic plane 
at intermediate to faint X-ray fluxes. A key question is whether
a reasonable number of such sources have the characteristics of 
CVs, consistent with the hypothesis that CVs contribute significantly
to the GRXE. 

The remaining sections of this paper are organised as follows. In the 
next section we discuss how we defined our source sample and extracted
X-ray spectral data for each source from the {\xmmn} science data 
archive. To carry out meaningful spectral analysis, we have divided our
sample of sources into 5 subgroups based on the source spectral hardness 
and stacked (\ie co-added) the individual source spectra within 
each subgroup (\S\ref{sec_3}). In
\S\ref{sec_4}, we analyse and compare
the stacked spectra for the different subgroups. We go on
to investigate the incidence of longer wavelength counterparts by
cross-correlating our X-ray source positions with 
NIR catalogues (\S\ref{sec_5}).  In section
\S\ref{sec_6} we discuss our results in the context of the
likely contribution of magnetic CVs to the GRXE.  
Finally we briefly summarize our main conclusions.


\section{The source sample and data reduction}\label{sec_2}

In Paper I we discuss the selection and properties of a sample of 2204
serendipitous X-ray sources drawn from 116 \xmmn observations with
pointings in the Galactic plane towards the central quadrant of the
Galaxy. Here we focus on a subset of sources from this original sample
with a view to extending the spectral analysis beyond the hardness
ratio and 'band index' considerations of Paper I.

The selection of objects for this new study was based on the following
requirements: (i) the source was detected in the EPIC pn camera in an
observation in which the medium filter was deployed; (ii) there were a
nominal 200-1200 pn counts associated with the source (more
specifically the product of the 0.5--12~keV pn count-rate times the
effective source exposure time was in the quoted range); (iii) if
there were multiple observations of the same source, the observation
with the longest exposure time and/or better quality data was
selected. After applying these criteria, the sample reduced to 138
sources drawn from 63 different \xmmn observations (hereafter we refer
to this set of 138 sources as the {\it current} sample).

\begin{figure}
\centering
\includegraphics[width=5cm,angle=-90]{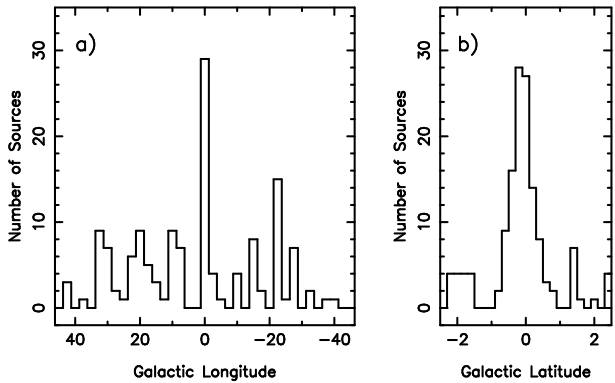}
\linespread{1}
\caption{(a) The distribution in Galactic longitude of the sources
comprising the current sample. (b) The same in Galactic latitude.}
\label{fig_1}
\end{figure}

Brief details of the sources comprising the current sample are given
in the Appendix. The distribution in Galactic longitude and latitude
is shown in Fig. \ref{fig_1}. The bulk of the sources are located
within $\pm 1^{\circ}$ of the Galactic plane and roughly 20 per cent
lie within $\approx 1^{\circ}$ of the Galactic Centre. 

\begin{figure}
\centering
\includegraphics[width=8cm,angle=-90]{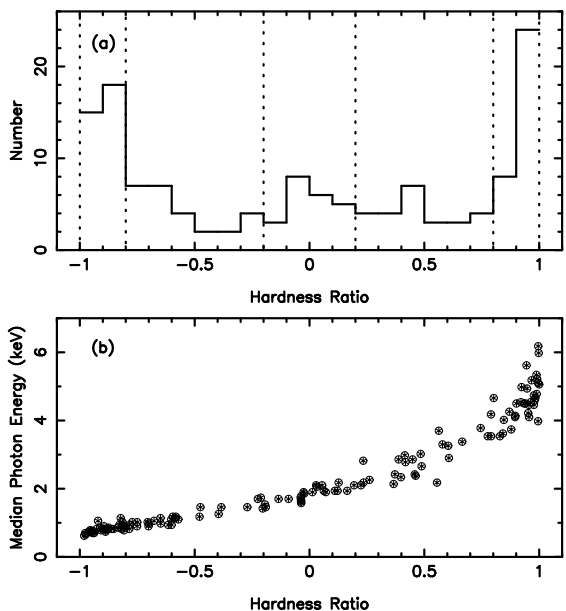}
\linespread{1}
\caption{(a) The distribution of the broad-band (0.5--2~keV : 2--12~keV)
hardness ratio within the current sample. The vertical dotted lines
define the ranges used to select the 5 source subgroups 
(see Table~\ref{table_1}). (b) The median photon energy (keV)
in the source spectrum plotted versus the broad-band hardness ratio
for each source.}
\label{fig_2}
\end{figure}

In Paper I we
categorised sources as spectrally hard or soft depending on whether
their broad-band hardness ratio, HR\footnote{More specifically, 
HR = (H-S)/(H+S), where H
corresponds to the 2XMMi Bands 4 and 5 (2--12~keV) and S to Bands 2
and 3 (0.5--2~keV) - see Paper I for details.}, was positive or
negative. In keeping with the
parent sample, the current sample splits fairly evenly into the soft
and hard source categories as illustrated in Fig. \ref{fig_2}(a). 
An alternative approach to the use of hardness ratios to
classify the spectral properties of X-ray sources with limited
count statistics is to employ `quantile' analysis (\eg
\citealt{hong04}) and, in that context, Fig. \ref{fig_2}(b) compares the 
{\it median} photon energy in the (background-subtracted) source spectrum,
$E_{50\%}$, with the corresponding value of HR for each source. For the
current sample, $E_{50\%}$ ranges from 0.5 keV at the soft
extreme up to to a maximum of $\approx 6$ keV for the hardest
sources.


Given that the source selection is based on a total count criterion,
there is a wide spread in the count rates of the selected objects.
For the great majority of the sources the count rate is in the 
range 5--100 pn count ks$^{-1}$ in either or both of the soft and hard
bands (cf. fig.~5 in Paper I). For the soft sources, 
the flux range sampled is  0.1--2 $\times 10^{-13}$  
\ergscm (0.5--2~keV)  and for the hard sources,  0.6--12
$\times 10^{-13} $\ergscm  (2--10~keV).

The present analysis is based on spectral data solely from the EPIC pn
camera. We extracted the pn spectrum of each source using the \xmmn
Science Analysis System (SAS) 
and standard techniques.
In brief, each dataset was reprocessed in order to apply the most
recent calibration and screened for high background flares by applying
a cut when the full-field count rate in the 10--12~keV band exceeded 0.4~ct
s$^{-1}$. Only single or double X-ray events were chosen with
\textsc{pattern} = 0--4 and with \textsc{flag} = 0 (the latter
excluding events close to chip gaps and bad pixels/columns). The
source spectra were extracted within a circular region of radius r =
35 arcsec (= 700 pixels) for most of the sources. However, for 13
sources the radius of the source extraction region was calculated
using the \textsc{SAS} task \textsc{region} in order to minimise
possible contamination from nearby sources. The background was 
taken from a circular extraction region with a radius of r = 120
arcsec (= 2600 pixels); in general, the background region was
positioned on the same CCD chip as the source (or occasionally on an
adjacent chip). Finally, the redistribution matrix files (RMFs) and
the ancillary response files (ARFs) were created using \textsc{rmfgen}
and \textsc{arfgen} for each individual source spectrum.


\section{Stacking of the source spectra}\label{sec_3}
After background subtraction, the net number of counts actually
recorded for each source was typically in the range 100--800 (see
Table~\ref{table_a1}). The measured counts were generally
reduced compared
to the prediction used in source selection due to two effects; the
loss of signal lying in the wings of the PSF outside of the source
extraction circle and the use of a more stringent threshold for the
data filtering than employed in the production of the 2XMMi catalogue.


\begin{table}
\small
\centering
\linespread{1}
\caption{The designation and hardness ratio range for the 5 source
subgroups. The number of sources contributing and the total counts per
subgroup, net of the background, are also quoted.}
\begin{tabular}{cccc}
\\
\hline
\hline
Subgroup & Hardness & Number of & Total net \\ 
& ratio & sources & counts \\
\hline
H & 1.0--0.8 & 32 & 8221 \\
MH & 0.8--0.2 & 25 & 6729 \\
MD & 0.2--(-0.2) & 22 & 7667 \\   
MS & -0.2--(-0.8) & 26 & 6906 \\
S & -0.8--(-1.0) & 33 & 11449 \\ 
\hline
\hline
\label{table_1}
\end{tabular}
\end{table}


Since individually the sources have insufficient counts for detailed
spectral analysis (although see \S\ref{sec_42}), 
the full sample of 138 sources
was divided into 5 subgroups (H, MH, MD, MS and S) based on the
source spectral hardness (see Fig.~\ref{fig_2} and 
Table~\ref{table_1}). Within each of
these five subgroups the source spectra were stacked. After applying
a scaling so as to match the area of the background region to that of
the source extraction region, the associated background spectra were
similarly co-added. The RMF and ARF files of each individual source
spectrum were combined by using the FTOOL \textsc{marfrmf}
v.2.2.0. Thereafter, the individual output files from \textsc{marfrmf}
were combined in \textsc{addrmf} v.1.21 so as to give a total response
file applicable to the stacked spectra for each source
subgroup. Finally, the spectra were binned using the FTOOL
\textsc{grppha}; 8-channel binning (of the 4096 input PHA channels)
was employed in the case of the spectra pertaining to the H, MH and
MD subgroups, whereas for the MS and S subgroups, 4- and 8-channel binning
was applied to the PHA ranges 0--511 and 512--4095, respectively.

Source statistics for the five subgroups are given in
Table~\ref{table_1}. Summed over the full set of 138 sources, 40972 pn
counts were recorded in the 0.5--12~keV band, net of the background.

\begin{figure*}
\centering
\includegraphics[width=12cm,height=12cm,angle=-90]{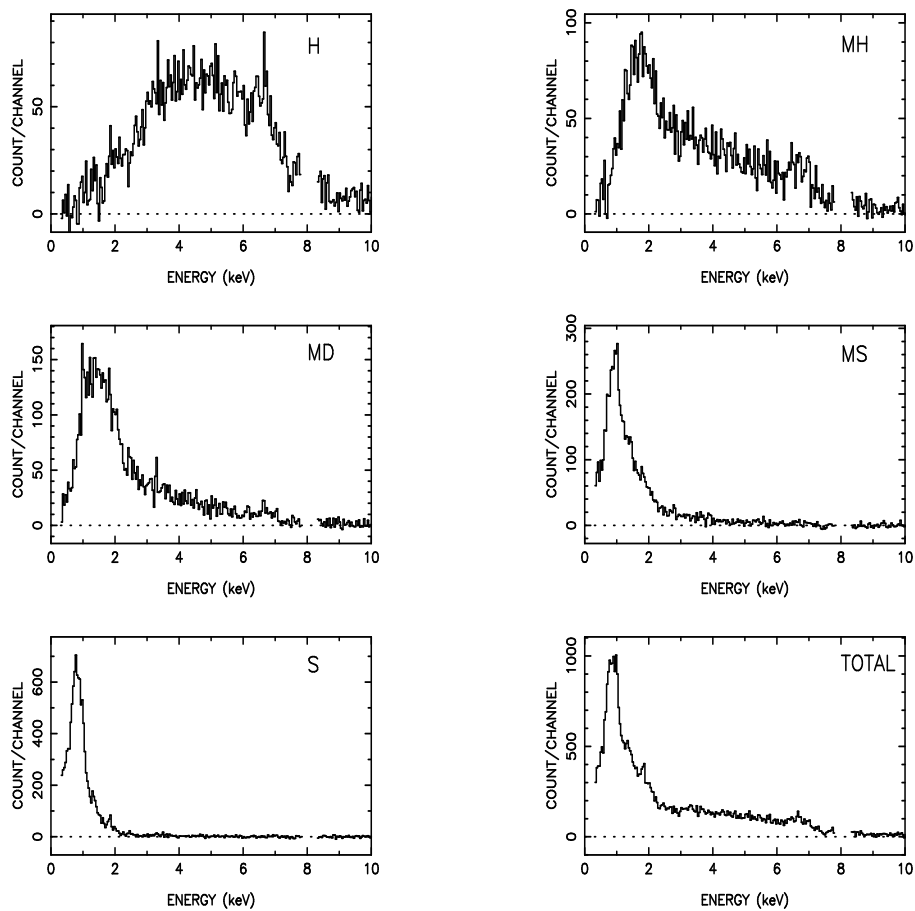}
\linespread{0.5}
\caption{The background-subtracted spectra for the 5 source subgroups
H, MH, MD, MS and S, together with the composite spectrum for all 138
sources (lower right corner). The data between 7.8 and 8.3~keV (i.e. the region of the K$\alpha$ Cu fluorescence lines in the
detector) have been excluded. The horizontal dotted line indicates
the level of zero net counts per channel.}
\label{fig_3}
\end{figure*}

The spectra of the five subgroups after background subtraction are
illustrated in Fig.~\ref{fig_3}, together with the composite
spectrum for the full sample. A very clear spectral softening trend is
evident from the H through to the S subgroups (rather as expected).
The spectra of subgroups H and MH show strong soft X-ray absorption
and also evidence for Fe-features in the 6--7~keV band. The MD subgroup
spectrum retains a hard tail out to $\sim$7~keV, which is barely present
in the spectrum of MS subgroup and completely absent from that of
the S subgroup.


\begin{table*}
\small
\centering
\linespread{1}
\caption{{\it Top section:} The best-fitting parameters from the spectral
fitting of an absorbed powerlaw plus emission lines model to the 
spectra of the H, MH and MD subgroups. The columns give the derived
column density, N$_{H}$, the photon index, $\Gamma$, the equivalent
width, EW, of the three Fe lines and  the reduced $\chi^{2}_{\nu}$ and
number of degrees of freedom $\nu$ for the fit.
{\it Lower section:} The best-fitting parameters from the spectral fits
to the spectra of the S, MS and MD subgroups, incorporating various 
combinations of thermal plasma and powerlaw components. The columns 
list N$_{H}$,  $\Gamma$,
two plasma temperatures kT$_{cool}$ and kT$_{hot}$, the derived metal
abundance Z$_{\odot}$ relative to solar. The
reduced $\chi^{2}_{\nu}$ and the number of degrees of freedom 
$\nu$ for the fit are also quoted.
}
\begin{tabular}{ccccccc}
\\
\hline
\hline
Group & N$_{H}$ & $\Gamma$ & EW(6.4 keV) & EW(6.7 keV) & EW(6.9 keV) & $\chi^{2}_{\nu}$/$\nu$ \\ 
& $\times$ 10$^{22}$ cm$^{-2}$ & & eV & eV & eV \\
\hline	
H & 8.3$^{+0.6}_{-0.5}$ & 1.55$^{+0.07}_{-0.07}$ & 94$^{+30}_{-30}$ & 149$^{+35}_{-34}$ & 55$^{+34}_{-33}$ & 1.16/517 \\
\\
MH & 1.27$^{+0.12}_{-0.10}$ & - & 152$^{+53}_{-53}$ & 172$^{+51}_{-51}$ & 192$^{+65}_{-65}$ & -\\
\\
MD & 0.34$^{+0.07}_{-0.06}$ & - & $ < 88$ & 187$^{+89}_{-91}$ & 56$^{+73}_{-56}$ & - \\
\\
Joint-EW & - & 1.54$^{+0.07}_{-0.07}$ & 89$^{+26}_{-25}$ & 170$^{+35}_{-32}$ & 81$^{+30}_{-29}$ & 1.20/523 \\
\hline
\hline
& N$_{H}$ &  $\Gamma$ & kT$_{cool}$ &  kT$_{hot}$ & Z$_{\odot}$ & $\chi^{2}_{\nu}$/$\nu$ \\
& $\times$ 10$^{22}$ cm$^{-2}$ &  & keV & keV & & \\
\hline
S & 0.057$^{+0.034}_{-0.022}$ & - & 0.51$^{+0.13}_{-0.12}$ &  1.01$^{+0.14}_{-0.06}$ & 0.13$^{+0.02}_{-0.02}$ & 1.17/159 \\
\\
MS & 0.18$^{+0.02}_{-0.02}$ &  -   &1.0 (fixed) & 3.6$^{+1.1}_{-0.7}$& 0.13 (fixed) & 0.85/160 \\
\\
MD & 0.87$^{+0.05}_{-0.05}$ & 1.55 (fixed) &  1.0 (fixed) & - & 0.13 (fixed) & 1.05/96 \\
\hline
\hline
\label{table_2}
\end{tabular}
\end{table*}


\section{X-ray spectral analysis}
\label{sec_4}

The analysis of the 5 stacked spectra was carried out using
\textsc{Xspec} version 12.8.1 \citep{arn96}. The spectral
fits were limited to the energy range 2--9~keV for the H subgroup and
to 1--9~keV for both the MH and MD subgroups. In each case, data
between 7.8 and
8.3~keV were excluded in order to eliminate the effect of background
subtraction residuals pertaining to the K$\alpha$ Cu fluorescence
lines in the detector background. For the MS and S subgroups, we used
the energy range 0.5--5~keV.


\subsection{The spectral models}
The X-ray spectra of the H, MH and MD subgroups were investigated using 
a spectral model comprising a powerlaw continuum plus three Gaussian
emission lines at fixed energies of 6.41, 6.68, 
and 6.96~keV. As noted earlier, these three lines correspond to K$\alpha$
emission from neutral (or near neutral) Fe-atoms, He-like Fe-ions and
H-like Fe-ions, respectively. The intrinsic widths of the three lines 
were fixed at $\sigma$ = 30~eV (\eg {\citealt{koy07};
\citealt{capelli12}). The emission components were subject to
a line-of-sight absorption column density, N$_H$, (using {\bf phabs}
in \textsc{Xspec}). A joint fit was employed, initially,
with only the powerlaw photon index, $\Gamma$, tied across the 
three datasets. The results
of this analysis are summarized in the first three lines of
Table~\ref{table_2}, where the line
strengths are quoted in terms of their equivalent width (EW) with 
respect to the underlying continuum. Here the errors are at 90\% 
confidence, except for the upper limit which is quoted at $3\sigma$.
 
From the results in Table~\ref{table_2}, it is evident that the simple
absorbed powerlaw plus lines model provides a reasonable description
of the data. The hardness ratio selection criterion used to define
the three source samples is reflected in the derived N$_{H}$ which varies 
from $\approx8~\times$ 10$^{22}$ cm$^{-2}$ for the H subgroup through to 
$ \approx3~\times$ 10$^{21}$ cm$^{-2}$ for the MD subgroup.  
The derived photon index, $\Gamma = 1.55 \pm 0.07$, demonstrates that
the sources contributing to the stacked spectra have, on average,
rather hard continuum spectra. When the powerlaw continuum is replaced by
a thermal bremstrahlung continuum, a correspondingly high temperature
is derived, kT $= 19^{+6}_{-4}$ keV, with only marginal changes in the
other spectral parameters and the $\chi^{2}$ of the fit.
  
The presence of a 6.7-keV Fe line is a feature common to the
H, MH and MD subgroups. There is also evidence for the 6.9-keV line
in the H and MH spectra and weakly in the MD spectrum.  Similarly
the 6.4~keV Fe line is detected in the spectra of the two hardest
subgroups, but not in the MD spectrum.
Fig.~\ref{fig_4} illustrates the form of the measured spectrum
and the corresponding best-fitting spectral model in the region
of the iron-line complex for the H subgroup.

In order to determine the line EWs averaged over the three datasets
we repeated the absorbed powerlaw plus Fe-lines spectral fitting with the
EWs of the three lines tied across the H, MH and MD spectra (in addition
to the photon index of the powerlaw continuum).  The result is reported
in  Table~\ref{table_2} on the line labelled {\it Joint-EW}.
In summary the average EW of the 6.7-keV line measured in the full
sample of sources with $HR > -0.2$ is $\approx 170$ eV, with
the 6.4-keV and 6.9-keV lines coming in at approximately
half this value.

A simple absorbed powerlaw model provided a poor description of the spectra
pertaining to the MS and S subgroups. Accordingly these spectra were
fitted using a two-temperature thermal plasma model,
which is a commonly used approximation for systems exhibiting a broad span
of emission measure versus temperature. 
In \textsc{Xspec} this was achieved by employing two {\bf apec}
components with a tied metal abundance Z$_{\odot}$.
The latter parameter was, nevertheless, allowed to vary (as a 
single scale factor applied to the standard solar abundance ratios 
defined by \citealt{anders89}).  The same foreground column density
was assumed to apply to both thermal components. 

The fitting of the spectrum of the softest subgroup resulted in 
temperatures close to 0.5 keV and 1.0 keV for the cooler
and hotter components, respectively - as reported in the lower section of
Table~\ref{table_2}. In this model the 1-keV component contributes 68\% of
the observed 0.5--2 keV flux.
The inferred metal abundance within thermal plasma is extremely subsolar, 
but this is most probably an artefact of the two-temperature 
spectral approximation, as has been noted by many authors 
(e.g. \citealt{str00}).
The combination of the low derived N$_{H}$ and a two-temperature description
of the emission spectrum are fully consistent with the underlying source
population being nearby, coronally-active stars. 

We next attempted to use the same emission model (kT values
fixed at 0.5 and 1 keV and Z$_{\odot}$ set to 0.13) to 
fit the spectrum of the MS dataset but failed to obtain a
satisfactory result. In fact, the presence of a (modest) hard tail within
this dataset requires a hotter thermal component, with the additional 
N$_{H}$ likely masking the soft emission characterized by the 0.5-keV 
component. The best fit obtained when the temperature of the latter
component was allowed to vary is reported in Table~\ref{table_2}.
In this case the 1-keV emission is the cooler component with
the hotter $\sim 3$ keV emission contributing
51\% of the observed 0.5-2 keV flux. The inferred 
column density for the MS subgroup is approximately three times that
of the S subgroup. If we (naively) interpret this in terms of the
MS sources being typically at three times the distance of those in the
S subgroup, then the implication is that the former
are on average an order of magnitude more X-ray luminous
(since the same flux threshold applies in the two cases). The
hotter emission of the MS subgroup might therefore reflect higher
levels of coronal-activity and the probable inclusion of some active
binaries.


\begin{figure}
\centering
\includegraphics[width=6cm,angle=-90]{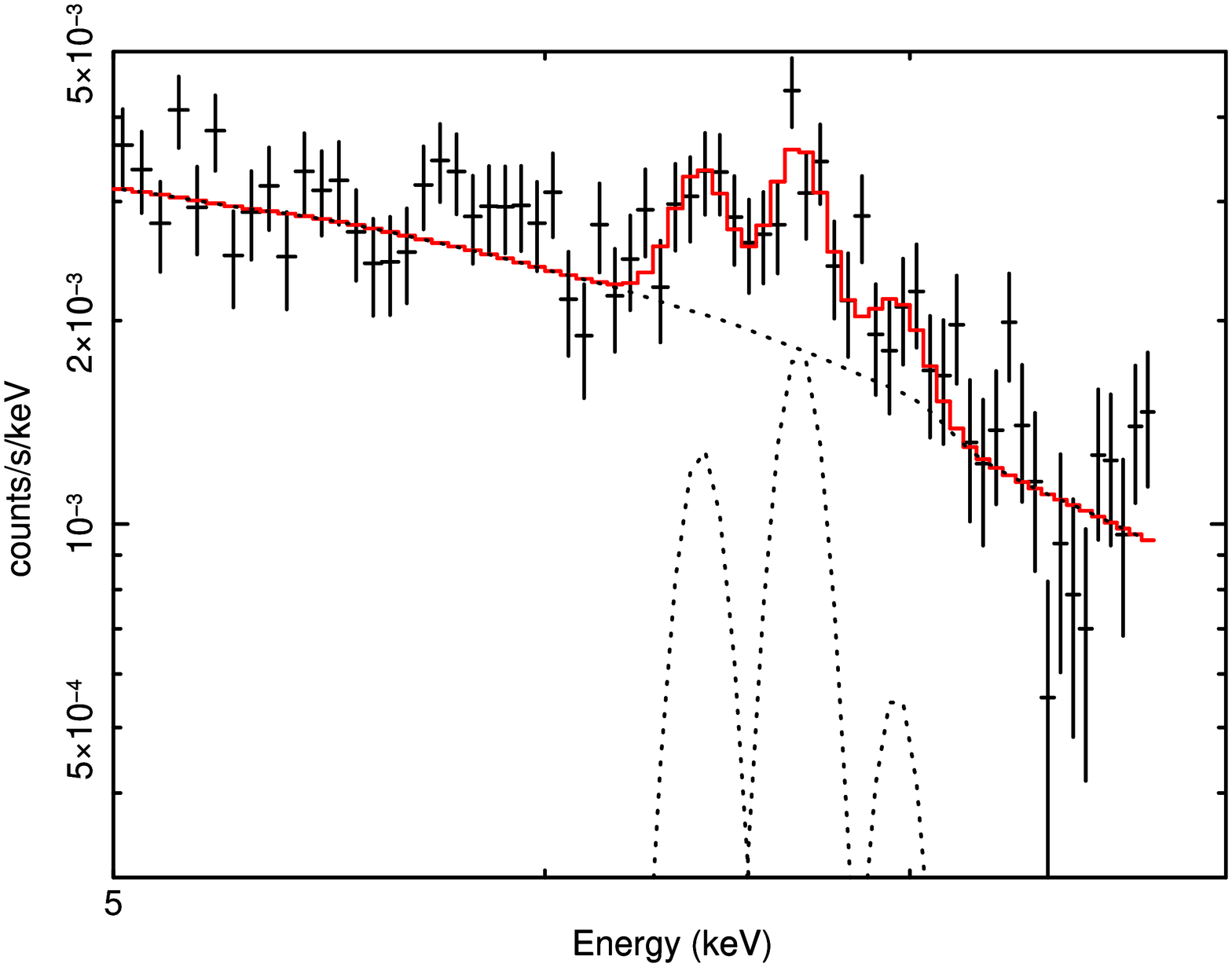}
\linespread{0.5}
\caption{The H subgroup spectrum around the iron-line complex.
The contribution to the best-fitting model of the powerlaw continuum 
and each of the Fe line components is shown as black dotted lines
with the combined fit shown in red.}
\label{fig_4}
\end{figure}


Given that coronally-active stars and binaries likely
dominate the MS and S subgroups, but certainly not the MH and H
subgroups (see Paper I and \S 5), it is safe to assume that the MD
subgroup includes both soft and hard population objects.
When we reconsider the spectral fitting of the MD subgroup
(in this case in the restricted 1-5 keV band), we find that
the inclusion of a 1-keV thermal plasma component (along with
the $\Gamma=1.55$ powerlaw) gives a significant improvement
to the fit (as measured by {\bf ftest} routine in \textsc{Xspec}).
The best-fit parameters for this powerlaw plus thermal model are provided
as the last entry in Table~\ref{table_2}. In this model, the thermal
component contributes roughly 20\% of the 1--5 keV flux.  Thus the MD
subgroup does seem to comprise a mix of source populations.
We note, however, that the presence of a number of sources with intrinsically
soft spectra in the MD subgroup is unlikely to have an undue impact on
the Fe-line EW measurements, since such sources will contribute little
to either the lines or the underlying continuum in the 6-7 keV band. 


\subsection{Fe-line properties of the individual sources}
\label{sec_42}

\begin{figure*}
\centering
\includegraphics[width=15cm,angle=-90]{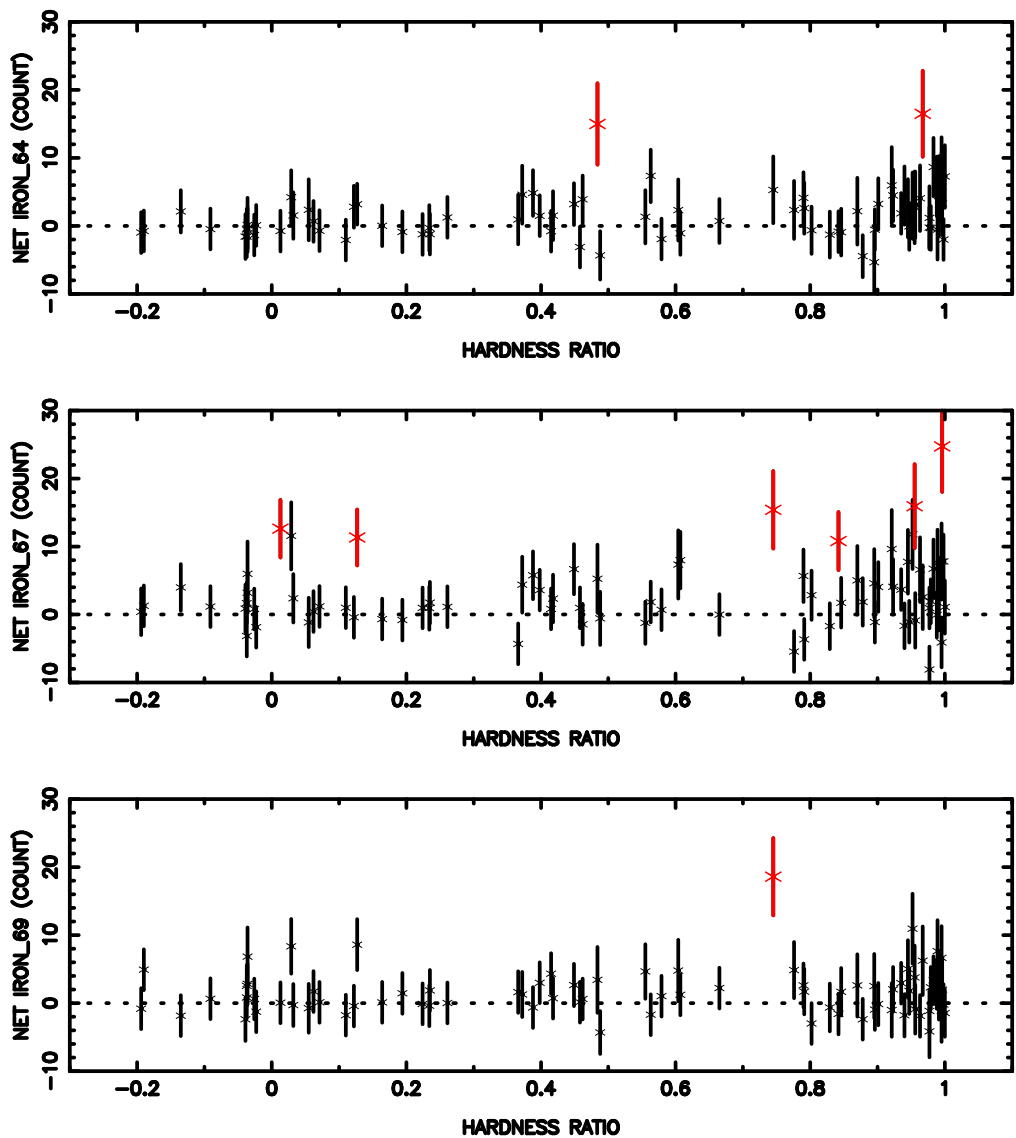}
\caption{The net counts of the 6.4-keV (top panel), 6.7-keV (middle) and
  6.9-keV (bottom panel) Fe lines versus the hardness ratio for each
  source. The red data points represent the sources with a
  significant (2.5$\sigma$) number of net counts in the line (see
  Table~\ref{table_3}).}
\label{fig_5}
\end{figure*}


\begin{table*}
\small
\centering
\linespread{1}
\caption{The 8 individual sources in which one or more of the 
Fe-lines (at 6.4, 6.7 and 6.9~keV ) was detected above 2.5$\sigma$.
The table lists the net count in the three lines for each source and
also quotes the EW if the line was detected above 2.5$\sigma$.}
\begin{tabular}{ccccccccc}
\\
\hline
\hline
Source & 212279 & 148374  & 168031 & 158770 & 166494 & 151081 & 217813 & 157004 \\
\hline
HR  & 0.01 &  0.13  & 0.48  & 0.75 & 0.84 &  0.95 & 0.97 & 1.00  \\
\\
Net cts (6.4 keV)  & -0.8 $\pm$ 3.0 & 3.2 $\pm$ 3.0 & 15.0 $\pm$ 6.0 & 5.3 $\pm$ 4.9 & -0.8 $\pm$ 3.0 & 2.7 $\pm$ 5.3 & 16.5 $\pm$ 6.3 & 3.3 $\pm$ 4.7 \\ 
EW (eV) & -- & --  & 674 $\pm$ 269 & -- & --  & -- & 244 $\pm$ 93 & -- \\
\\
Net cts (6.7 keV) & 12.6 $\pm$ 4.2 & 11.3 $\pm$ 4.1 & 5.3 $\pm$ 5.0  & 15.4 $\pm$ 5.7 & 10.8 $\pm$ 4.3 & 15.9 $\pm$ 6.2 & 2.6 $\pm$ 4.7 & 24.7 $\pm$ 6.7  \\
EW (eV) & 1218 $\pm$ 409 & 2945 $\pm$ 1066  & -- & 508 $\pm$ 188 & 1210 $\pm$ 477 & 329 $\pm$ 127 & -- & 635 $\pm$ 171 \\
\\
Net cts (6.9 keV) & 0.1 $\pm$ 3.0 & 8.6 $\pm$ 3.7 & 3.4 $\pm$ 4.8  & 18.6 $\pm$ 5.7 & -1.6 $\pm$ 3.0  & 3.8 $\pm$ 4.7 & 6.2 $\pm$ 5.0 & 2.0 $\pm$ 4.4 \\
EW (eV) & -- & --  & -- & 686 $\pm$ 210 & -- & --   & -- & -- \\ 
\hline
\hline
\label{table_3}
\end{tabular}
\end{table*}


\begin{figure*}
\centering
\includegraphics[width=6cm,angle=-90]{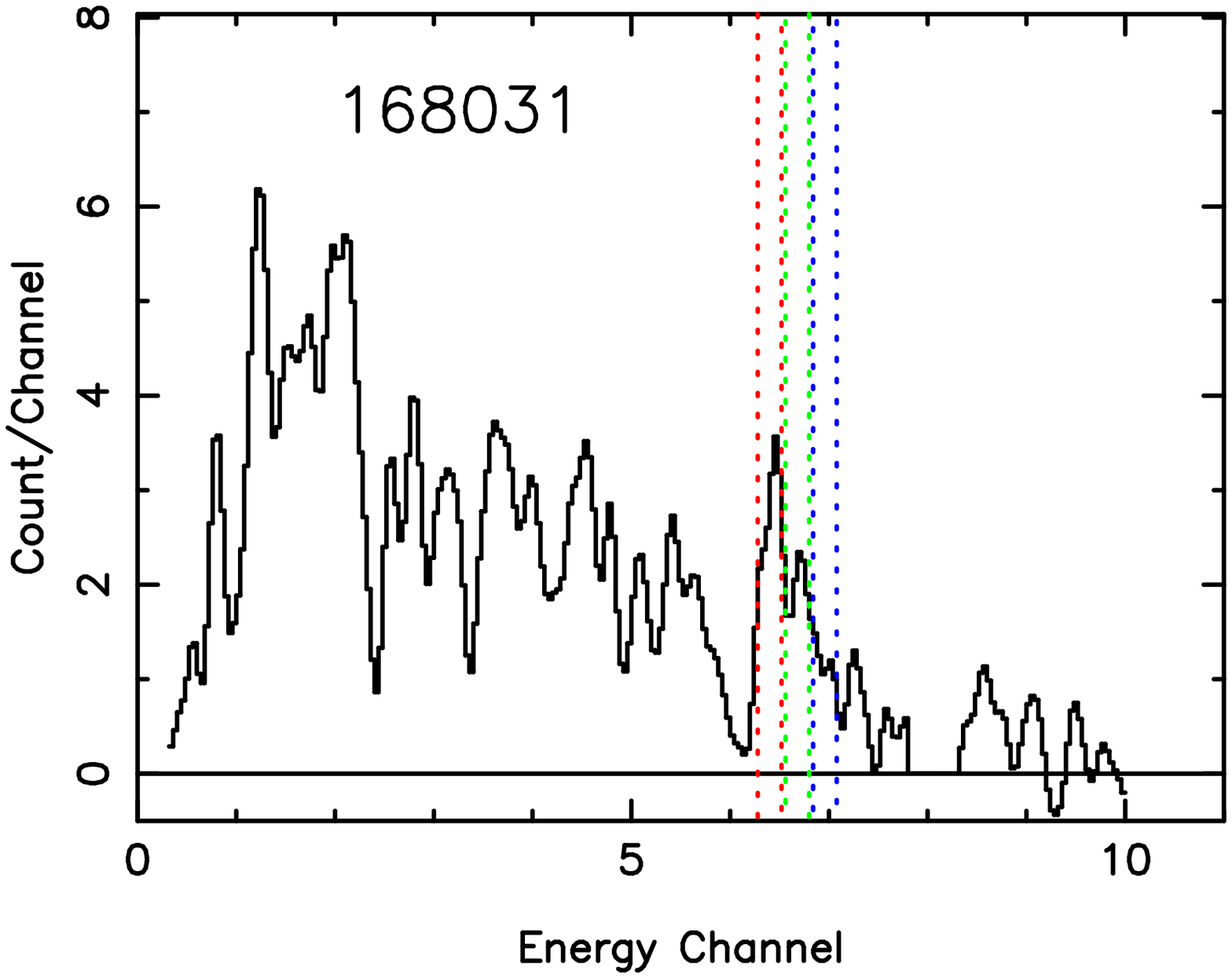}
\includegraphics[width=6cm,angle=-90]{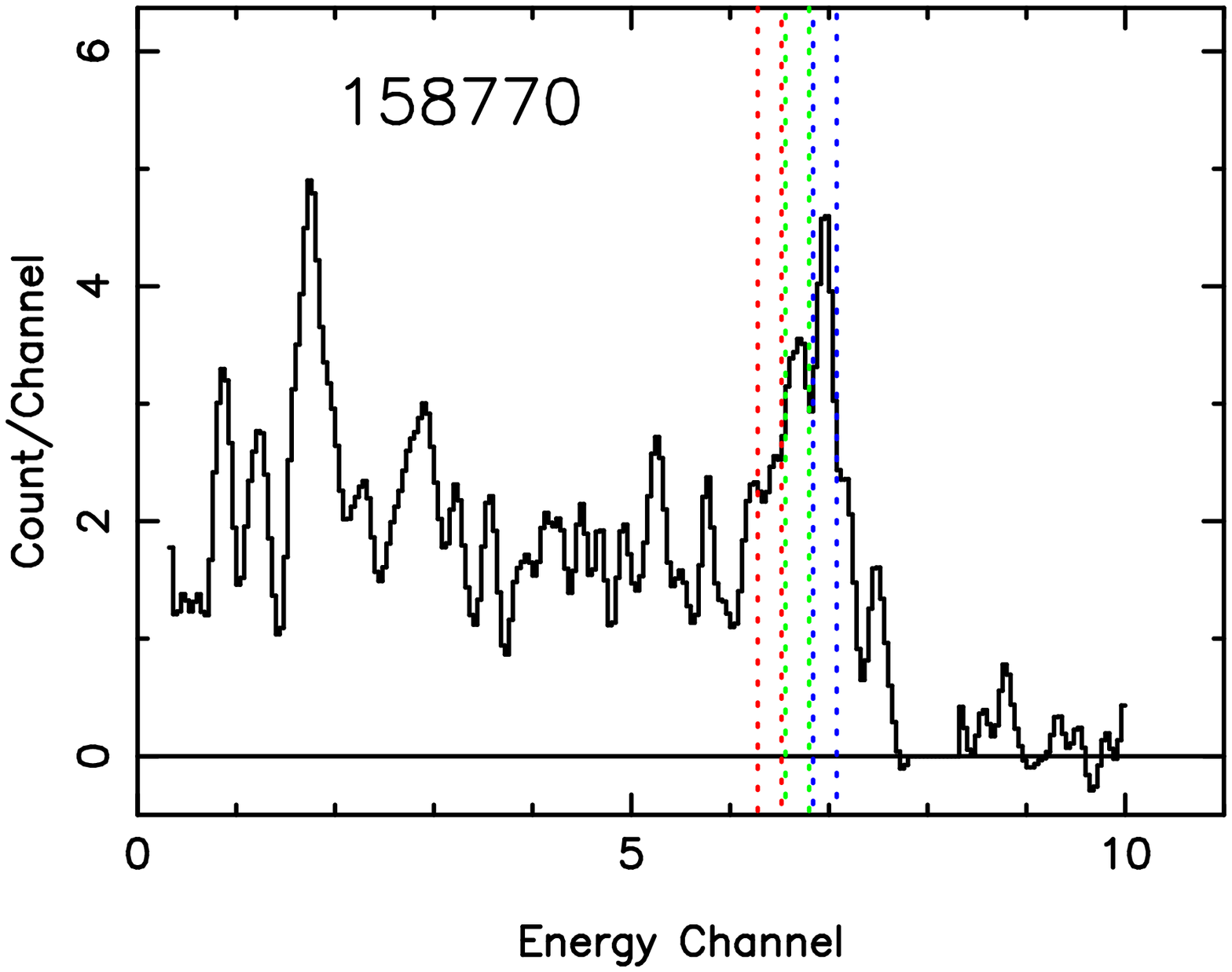}
\includegraphics[width=6cm,angle=-90]{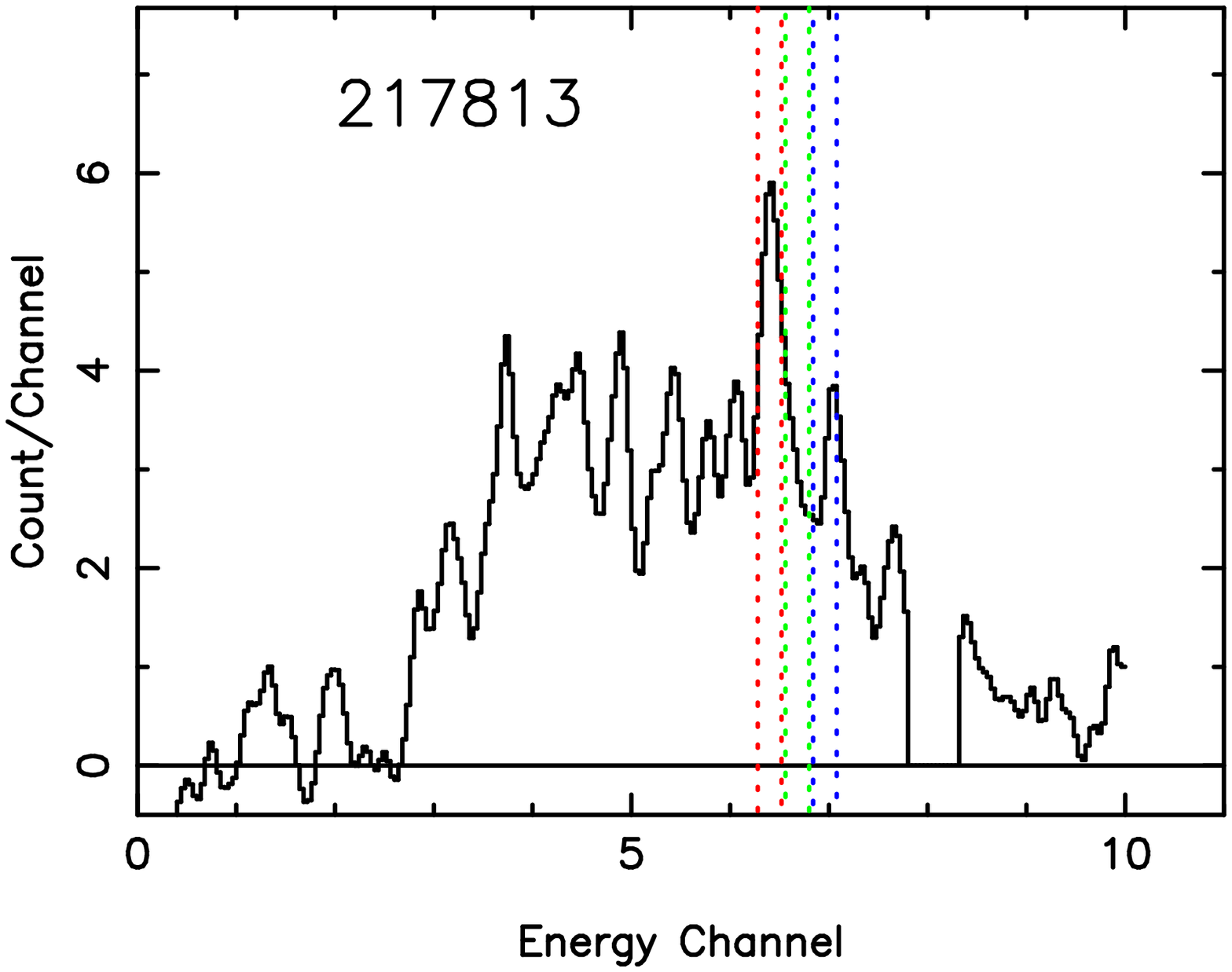}
\includegraphics[width=6cm,angle=-90]{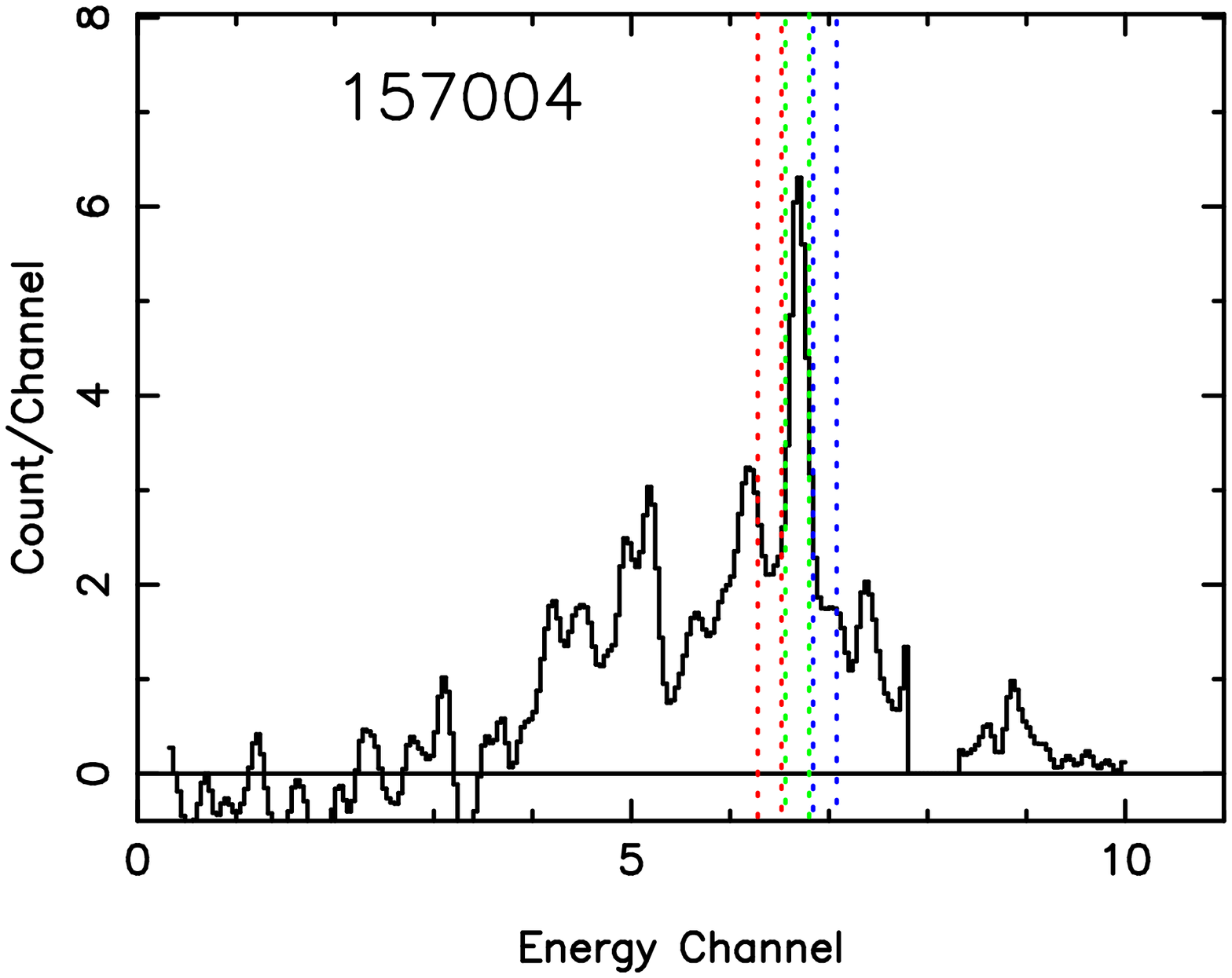}
\caption{The measured source spectrum net of background for four
sources. The spectral data have been grouped into 40~eV channels and
smoothed with a Gaussian line spread function of width $\sigma$=60~eV.
The vertical dashed lines indicate the 6.4 (red), 6.7 (green) and
6.9~keV (blue) bandpasses. The data between 7.8 and 8.3~keV ({\it
i.e.,} the region of the K$\alpha$ Cu fluorescence lines in the
detector) have been excluded. {\it Top panels:} Two sources from the
MH subgroup with detectable lines at 6.4 keV (168031) and 
6.7 and 6.9 keV (158770).  {\it Bottom panels:} H subgroup sources
with line features at 6.4 keV (217813) and 6.7 keV (157004).}
\label{fig_6}
\end{figure*}


In order to investigate the contribution of individual sources to the
Fe-line emission in the H, MH and MD subgroups, we extracted the
counts recorded for each source in a narrow bandpass (of width 240 eV)
centred on each Fe line. The level of the underlying continuum
appropriate to each measurement was also estimated based on a linear
interpolation of the counts recorded in two `continuum' bands (each
of width 600~eV) centred at 5.94 and 7.54~keV. Fig.~\ref{fig_5}
shows a plot of the net line counts in the 6.4-, 6.7- and 6.9-keV lines
versus the hardness ratio for the full set of sources that comprise
the three subgroups. In this figure the data points plotted in red
correspond to those sources for which the line measurement represents
a 'detection' at or above the 2.5$\sigma$ significance level.
The number of sources with directly detectable lines at 6.4, 6.7 and 
6.9~keV is 2, 6 and 1, respectively, with just one source exhibiting
more than a single line (158770 - with twin
features at 6.7 and 6.9 keV). A summary of the Fe-line
properties of these 8 sources is provided in Table~\ref{fig_6}.

In broad terms, Fig.~\ref{fig_5} confirms the results of the
spectral fitting of the source subgroups. In the case of the 6.7-keV
line, there is a positive bias in the EW measurements, particularly
for the sources within the H and MH categories (HR $>$
0.2). The 6.4 and 6.9~keV measurements also fit this description.

There is a suggestion that the relatively high line EWs measured in the
MH subgroup (see Table~\ref{table_2}) may be due to the influence of two
sources - source 168031 which has a relatively bright 6.4 keV
line and, as noted above, source 158770 which exhibits detectable
line features at both 6.7 and 6.9 keV. Fig.~\ref{fig_6}
illustrates the X-ray spectra of these two sources
in comparison to the spectra of two 'typical' sources drawn from
the H subgroup,
one with a detectable 6.4-keV line and the other with a 6.7-keV feature.
If we exclude the two sources identified above from the MH sample, the
impact is to reduce the EW values tabulated for the MH subgroup in
Table~\ref{table_2} by about 30\%; however, such an exclusion
would be somewhat arbitrary.


\section{NIR counterparts of the sources}
\label{sec_5}

We have searched for potential NIR counterparts of the
X-ray sources in the current sample by cross-matching the \xmmn source
positions with the Two Micron All Sky Survey 
(2MASS)\footnote{http://www.ipac.caltech.edu/2mass/}
(\citealt{cutri03}; \citealt{skrutskie06}).
The methodology for the cross-matching was essentially the same as that
reported in Paper I, where full details are provided.

We found that of the 59 sources in the S and MS subgroups, 53 
(90\%) have a bright ($K_{S} <$ 14) 2MASS star within a $3\sigma$ 
error radius of the X-ray position (i.e. within an
error circle with radius equal to 3 times the X-ray position error
quoted in the 2XMMi 'slim' catalogue{\footnote{In the case of multiple
observations of an X-ray source, the X-ray position and
position errors quoted in the 2XMM 'slim' catalogue is the weighted
average value across {\it all} the detections of the source.}). 
There are 3 further likely 2MASS counterparts if the error circles are
extended to $3.25\sigma$. The remaining 3 sources include one instance
where the 2MASS catalogue suffers from the confusion of a bright nearby
star and, similarly, two instances where the X-ray position may be
marginally offset due to the confusion of a nearby X-ray source.
In short, virtually all of the soft X-ray sources have plausible
NIR-bright counterparts, the majority of which are likely to be nearby, 
coronally-active stars and binaries (see Paper I).

The NIR associations for the harder spectral subgroups are much less
complete. For the MD subgroup, 13 sources out of 22 (59\%) have bright
($K_{S} <$ 14) 2MASS objects in a nominal $3\sigma$ X-ray error
circle, whereas for the MH and H subgroups the statistics are 10 out
of 25 (40\%) and 8 out of 32 (25\%), respectively.  The average
(area-weighted) $3\sigma$ error circle radius for the current sample
of sources is 2.5 arcsec. Utilising a set of positions offset from
each X-ray position, we find that, for these Galactic plane fields,
the corresponding chance coincidence rate with 2MASS stars brighter
than $K_{S} = 14$~ is 15\%. Fig. \ref{fig_7} summarises these
statistics.  The implication is that, at least within the MD and MH
subgroups, a sizeable minority of sources may have real
identifications with bright NIR stars.

We have plotted a NIR two-colour diagram for the 2MASS
stars contained within the X-ray error circles in  Fig. \ref{fig_8}
(but excluding sources with relatively poor photometry
for which the 2MASS Q flag $=$ U in one or more
bands). In this diagram the bulk
of the objects associated with X-ray sources in the S and MS 
categories fall on the locus of late-type main sequence stars with
relatively low reddening. In contrast, the stars linked to the
X-ray sources in the MD through to H subgroups are characterized
by an increasing degree of reddening.


\begin{figure}
\centering \includegraphics[width=6.3cm,angle=270]{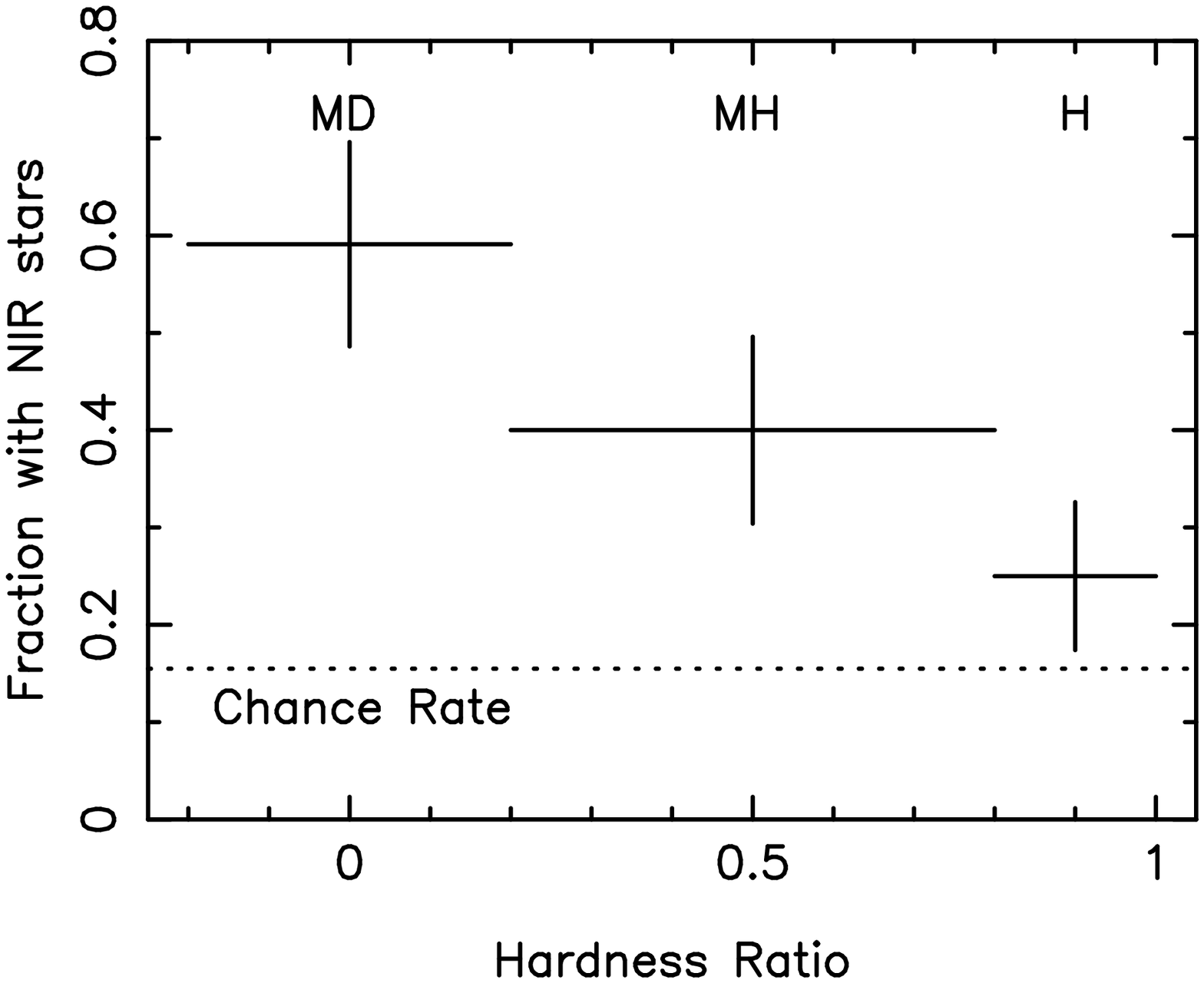}
\linespread{1}
\caption{The fraction of sources with a bright 2MASS star contained 
within the X-ray error circle for the MD, MH and H subgroups.
The chance rate is indicated by the dotted line. 
}
\label{fig_7}
\end{figure}



\begin{figure}
\centering \includegraphics[width=6.3cm,angle=270]{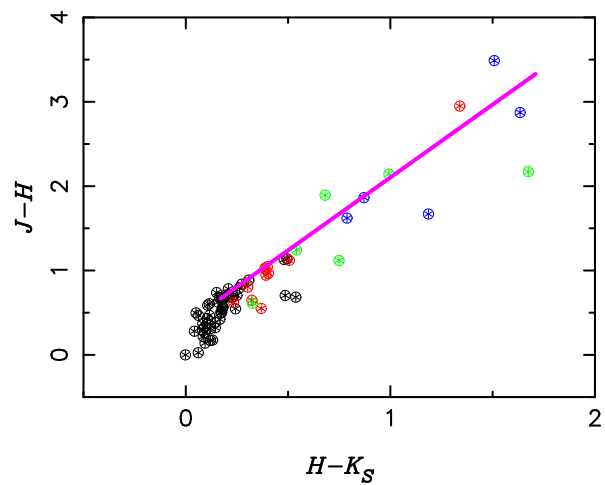}
\linespread{1}
\caption{The $H-K_S$ versus $J-H$  colour-colour diagram of the brightest
2MASS star contained within the X-ray error circle. X-ray sources 
in the S and MS subgroups are shown as the black symbols, 
whereas those in the MD, MH and H subgroups are plotted in red, green 
and  blue, respectively. The solid
diagonal line  illustrates the reddening vector corresponding to $A_{V} = 25$,
($A_{J} = 7.0$; $A_{H} = 4.3$; $A_{K} = 2.8$), from a starting point
representative of  the NIR colours of an unreddened MOV star.}
\label{fig_8}
\end{figure}



\begin{figure*}
\centering \includegraphics[width=10.0cm,angle=270]{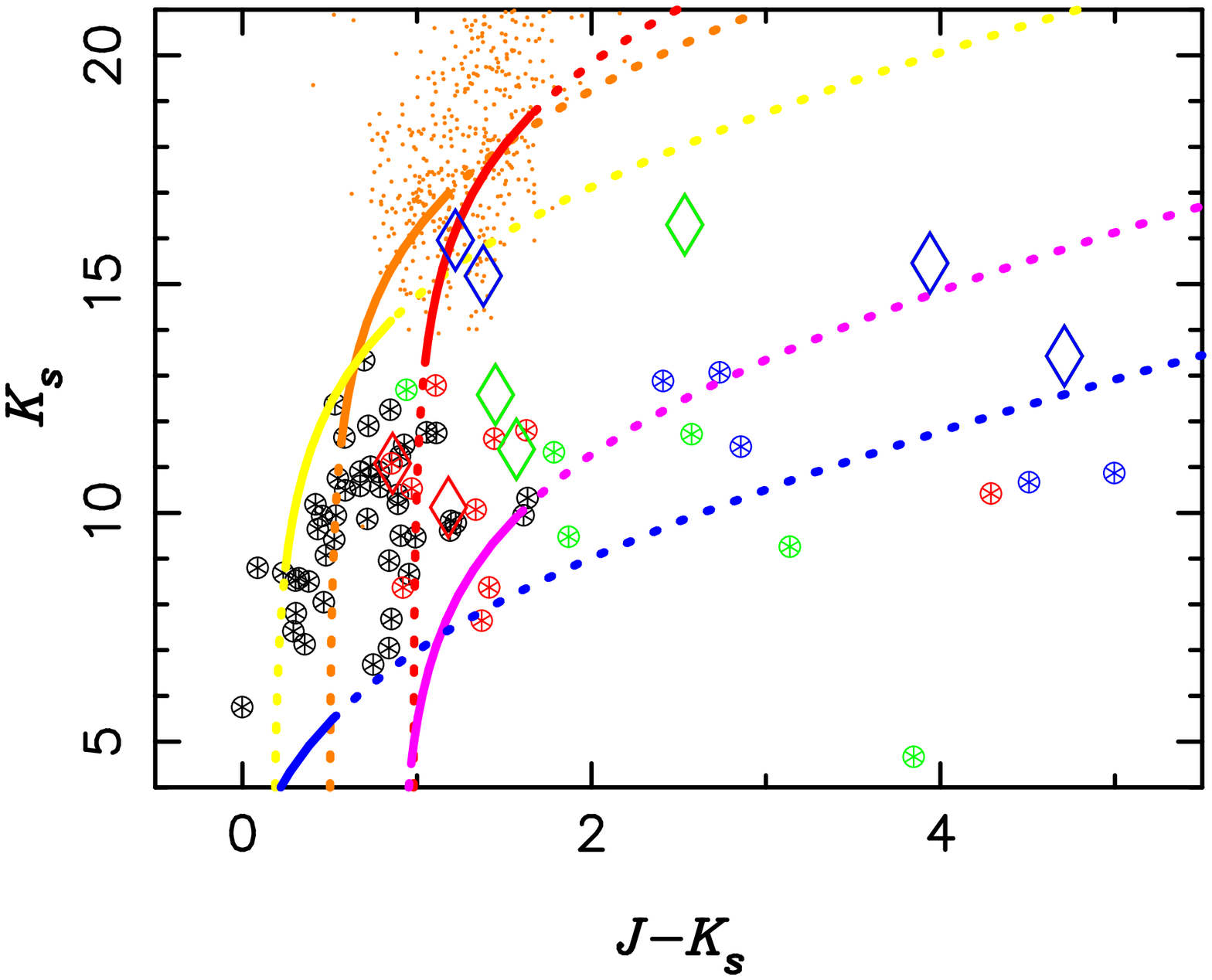}
\linespread{1}
\caption{The $J-K_S$ versus $K_S$ colour-magnitude diagram of the
brightest NIR stars contained within the X-ray error circle. 
The results from the comparison with 2MASS are shown as the
circular symbols with colour coding indicative of the X-ray subgroup
as follows - S and MS (black), MD (red), MH (green) or H (blue).
The diamond symbols show the same information when the study is extended
to UKIDSS for a subset of the sources - see the text. 
The cluster of brown points represents known CVs when 
their colours and magnitudes are translated to a standard
distance of 2 kpc.
The curves illustrate the tracks of different
types of star as the distance is varied from 1 pc - 20 kpc, assuming that
$A_{V}$ in the Galactic plane increases at the rate of 2 mag kpc$^{-1}$. 
The curves become solid lines for stellar distances in the range 
200 pc - 2 kpc, which corresponds to an X-ray luminosity
in the range 10$^{30-32}$ erg s$^{-1}$, assuming a limiting
X-ray flux of $2 \times 10^{-13}$ \ergscm (2--10 keV). 
{\it Brown curve:} a luminous CV with M$_{K}$ = 5.0
and intrinsic $J-K_S$ = 0.5; {\it Yellow curve:} a FOV dwarf star with
M$_{K}$ = 2.25, $J-K_S$ = 0.18; {\it Red curve:} a
M6V dwarf with M$_{K}$ = 6.74,
$J-K_S$ = 0.98; {\it Purple curve:} a K3III giant with M$_{K}$ = -2.03,
$J-K_S$ = 0.90;  {\it Blue curve:} a BOI supergiant with 
M$_{K}$ = -6.44, $J-K_S$ = -0.15.
}
\label{fig_9}
\end{figure*}


We have also plotted a colour-magnitude diagram for the same set of
NIR objects in Fig.\ref{fig_9}  (as the circular symbols). 
This figure also shows the tracks of
several different types of stellar objects in the colour-magnitude
plane assuming that the visual absorption, $A_{V}$, 
in the Galactic plane increases 
at the rate of 2 mag kpc$^{-1}$ (equivalent to $A_{J} = 0.56$ mag kpc$^{-1}$; 
$A_{K} = 0.224$ mag kpc$^{-1}$). We use a compilation of NIR magnitudes
for CVs \citep{ak07, ak08} to estimate the
absolute magnitude and intrinsic colour typical of a CV. Similarly, the 
NIR properties of the dwarf, giant and supergiant stars 
are taken from \citet{covey07}. 
This figure illustrates rather clearly that if the association
of an X-ray source with a {\it highly reddened}, but {\it relatively bright}
NIR star (with, say, $J-K_{S}>$ 1.0 and $K_{S} < $ 14) is real, then
it is unlikely that we are dealing with either a distant luminous CV or 
a very coronally-active dwarf (F-M) star. The basic point is that if we
interpret a high degree of reddening as indicative of a relatively
large distance (i.e. assuming that the reddening is not a feature
of the very local environment of the source) then rather faint
magnitudes are implied for dwarf stars. It is much more tenable that
the association is with either a late-type giant or a bright
supergiant (e.g. \citealt{mau09,mau10}; \citealt{motch10}; \citealt{nebot13}). 
The former category includes RS CVn binaries containing a 
G- or K-type giant twinned with a sub-giant or main sequence companion,
which are characterized by enhanced, high-temperature coronal emission
driven by tidal interactions \citep[e.g.][]{agr81}. 
If, for example, we observe an RS CVn system containing a K3III giant
at $K_{S} = 10$, then the implied distance is $\approx 2$~kpc. For a 
limiting flux in the hard band of, say, $2 \times 10^{-13}$ 
\ergscm (2--10 keV), the implied X-ray luminosity, L$_X$, is $
\approx 10^{32} \rm~erg~s^{-1}$, which is plausible for
the most active systems. If the counterpart to the X-ray source
is a massive early-type star such as a Wolf-Rayet star or an 
OB-supergiant, then a significant X-ray flux above 2 keV may be produced
in shocked regions within an unstable wind or, in the case of massive
binaries, in the colliding winds. As is evident from Fig.\ref{fig_9}, 
in this case a $K_{S}$ magnitude of 10 or fainter
implies a hard-band X-ray luminosity considerably greater than
$10^{32} \rm~erg~s^{-1}$, which points to either a binary (or multiple) 
massive star system or the presence of an accreting high-mass X-ray binary 
(Paper I; \citealt{mau10}; \citealt{ander11} and references therein).

Fig.\ref{fig_9} also shows a cluster of points representative of
the colours and magnitudes of known CVs, when translated to
a common distance of 2 kpc, based on the compilation of CV
NIR magnitudes and distances reported by \citet{ak07, ak08}. As noted
above for a distance of 2 kpc, our hard band limiting flux
corresponds to an X-ray luminosity of $\approx 10^{32}$ \ergs.
Clearly, in order to identify such a population of sources
amongst our sample of hard X-ray sources, one would need to go 
to significantly fainter NIR magnitudes than are probed by 2MASS. 
Typically such conterparts would have $K_{S} > 15$, but be subject
to only a modest degree of interstellar reddening.
The problem, of course, is that in the Galactic plane
at these faint IR magnitudes, the chance incidence rate
of stars may become excessive, at least when using the typical
few arcsec error circles available from \xmmn.

By way of an experiment, we have selected a subset of the 
hard X-ray sources (within the MD, MH and H subgroups) which
have relatively small X-ray error circles ($3\sigma$ error
circle radius $\le 1.5$\arcsec) and  for which deep NIR imaging is
available from the United Kingdom Deep Sky Survey 
(UKIDSS)\footnote{http://www.ukidss.org/}\citep{law07}. Of the 18 X-ray
sources meeting this criterion, we have found possible NIR
associations in 9 cases. These are shown as the diamond symbols
in Fig.\ref{fig_9}.   Two of these NIR objects (both linked to
H subgroup sources) are located in the region of the diagram where
we might expect to find X-ray luminous CVs. Interestingly a search 
of the catalogues held at 
VizieR\footnote{http://vizier.u-strabg.fr/viz-bin/Vizier/} shows
that one of these objects is a known Intermediate Polar
(also known as SAX J1748.2-2808; \citealt{ritter03}; \citealt{sidoli06};
\citealt{nob09}).  A VizieR search on the full sample of hard
X-ray sources also revealed a handful of other likely identifications,
including several high reddened sources near the Galactic Centre
previously studied by \citet{mau09}, two LMXB \citep{liu07}, a Be-star
HMXB \citep{ebi03} and a potential AM HER object \citep{motch10}. 

In summary, our cross-correlation with NIR catalogues demonstrates 
that virtually all the soft X-ray sources in our sample (comprising
the S and MS subgroups) can be associated with a bright late-type 
stellar counterpart possessing an active stellar corona.
The nature of the Galactic population or populations underlying the hard
X-ray sources (comprising subgroups H, MH and MD) is less clear-cut,
although highly coronally-active systems such as RSCVn binaries,
wind dominated objects such as Wolf-Rayet stars and OB-supergiants
and accretion-powered systems ranging from CVs, through to LMXB 
and HMXBs, all contribute to the mix.  Extragalactic
interlopers, predominantly Active Galactic Nuclei (AGN) 
seen through the high column density of the Galactic plane
may also represent a significant (maybe 50\%) fraction of
the hardest subgroup of sources (Paper I; \citealt{hands04}).
However, our analysis does provide a hint that  an appropriate
combination of sub-arcsec X-ray positions and deep high resolution
NIR imaging has the potential to reveal an increasing number of CV
counterparts, at least in surveys reaching intermediate rather than 
very deep X-ray flux thresholds.


\section{Discussion}
\label{sec_6}

Currently, the leading candidate for the unresolved source population 
contributing to the GRXE is {\it magnetic} CVs, nearby examples of
which typically have 2--10 keV X-ray luminosities in the range
from $10^{31}$ -- $10^{34}$ erg s$^{-1}$ (e.g. \citealt{ezuka99};
\citealt{saz06}; \citealt{rev08}). In principle this
includes both polars (where the spin and orbit are synchronized
or nearly synchronized) and Intermediate Polars (IPs)
(where the orbital period is much longer than the white dwarf spin
period).  However, the plasma temperatures of polars are somewhat lower than
in IPs due to  enhanced cyclotron cooling  (e.g. \citealt{cropper98})
and it is likely that IPs contribute most to the
hard X-ray ($>$ 5 keV) volume emissivity.

In magnetic CVs, an accretion shock is generated above the
white dwarf surface, which heats the accreting material to 
temperatures kT $> 15$ keV. The resulting highly-ionised, 
optically-thin plasma cools in the post-shock flow and, eventually, 
settles onto the white dwarf surface through an accretion column. 
The resulting X-ray spectrum comprises a complex blend of contributions
generated at differing temperatures, densities and optical depths
(e.g. \citealt{cropper99}; \citealt{yuasa10}). 
As well as a hard continuum, the X-ray spectra of magnetic CVs
are characterized by a complex
of iron-K lines including strong He-like, H-like and fluorescent
components, the latter arising from the illumination
of the surface of the white dwarf and/or the outer regions
of the accreting flow by the hard continuum.

In a study of 20 magnetic CVs observed by {\it ASCA}, \citet{ezuka99}
found the characteristic temperature of the underlying
hard continuum to be typically kT $\sim 20$ keV. 
Although with a large source-to-source scatter, the average EW of the
6.4-keV, 6.7-keV and 6.9-keV lines across this CV sample were
roughly $\sim 100$ eV, $\sim 170$ eV and $ \sim 100$ eV, respectively. 
Similar average EWs were reported in the parallel
study by \citet{hel98}.
More recently, \citet{hel04} investigated the X-ray spectra
of 5 magnetic CVs
observed at high spectral resolution with the {\it Chandra} High-energy
Transmission Grating (HETG). Within this limited sample the average EWs
were $\sim 120$ eV, $\sim 160$ eV and $\sim 110$ eV for the 6.4-keV, 6.7-keV
and 6.9-keV lines, respectively. Finally, \citet{bern12} report the detection
of intense Fe-K$\alpha$ line emission in the X-ray spectra of 9 recently
identified IPs and quote the EW of the 6.4-keV fluorescent line, in all
cases, to be in the range 130-220 eV.

In our earlier analysis (\S\ref{sec_4}), we found that the
emission spectra of the three hardest subgroups could be represented as
a hard continuum plus Fe-lines.  We modelled the hard continuum
as a powerlaw of photon index $\Gamma =1.55$ but obtained equivalent results
when the powerlaw was replaced by a thermal bremstrallung continuum
with kT $\approx 20$ keV.  The average EW of the He-like Fe-line at 
6.7 keV was $170^{+35}_{-32}$ eV.  Similarly, the EWs of the 6.4-keV and
6.9-keV Fe-lines were $89^{+26}_{-25}$ eV and $81^{+30}_{-29}$ eV, respectively, 
\ie roughly half that of the 6.7-keV line.  Evidently, these spectral
characteristics are fully consistent with those of known magnetic CVs.

The inference is that a sizeable fraction of our current hard band sample
may be magnetic CVs, albeit with counterparts which are in general too faint
to pick out reliably from available NIR catalogues. However, it is necessary to
consider what impact non-CV interlopers might have on the measurements
deriving from the stacked (\ie sample-averaged) spectra.  For example,
we noted in \S\ref{sec_5} that, based on source count estimates,
AGN might contribute up to 50\% of the hardest subgroup of sources.
Compton-thin AGN are characterized by powerlaw continuum spectra
with $\Gamma \approx 1.7$ plus a prominent Fe fluorescence line 
with an EW typically in the range 20-120 eV (see \citealt{fuk11}; 
\citealt{chaud12} and references therein).
Clearly any dilution of the CV contribution by AGNs would tend to
reduce the EW of the 6.7-keV and 6.9-keV lines, whilst having relatively
little impact on the 6.4-keV measurements. There is perhaps slight
evidence for this from a comparison of the results from the
three hardest subgroups (on the presumption that the high
column density through the Galactic disc places AGN preferentially
in the hardest subgroup); however it seems unlikely that this
has introduced a strong overall bias in our Fe-line EW measurements.

Having demonstrated, on spectral grounds, that magnetic CVs very
likely represent a significant Galactic population at the intermediate to
faint X-ray fluxes encompassed by our \xmmn source sample, what are the
implications for the origin of the GRXE?  
In a recent study of the distribution of K-shell line emission along the
Galactic plane observed by \suzaku,
\citet{uch13} find that the average EW of the 6.4-keV, 6.7-keV and 6.9-keV
lines associated with the GRXE are $\approx 110$ eV, $\approx 490$ eV and 
$\approx 110$ eV (see also \citealt{ebi08}; \citealt{yam09}).
As noted by \citet{uch13} and other authors,  although the 6.4-keV
and 6.9-keV line EWs measured for the GRXE match those of magnetic
CVs, it appears that the EW of the 6.7-keV line is in a factor of 2-3
times higher than is typical of such systems.

Of course, our present measurements pertain to sources
with X-ray fluxes down to a limiting threshold of roughly
2 $\times 10^{-13} $\ergscm  (2--10~keV), by which point less than
$\sim 10\%$ of GRXE intensity has been resolved (\citealt{hands04};
\citealt{ebi05}). In fact it has been pointed out
that in order to resolve 90\% of the GRXE, a sensitivity limit
of $\sim 10^{-16}$ \ergscm  (2--10~keV) will be needed, encompassing
sources with luminosities as low as $10^{29-30}$ \ergs ~at the Galactic
Centre distance (\citealt{rev06, rev09, rev11}; \citealt{morihana13}). 

At these much fainter fluxes and luminosities, the balance across
the various X-ray emitting populations will very likely change
(e.g. \citealt{morihana13} - although for a counter argument 
see \citealt{ruiter06}; \citealt{hong09}; \citealt{hong12}). 
For example, non-magnetic CVs, which
comprise the majority of the local CV population, are well-established
sources of X-ray emission at the lower end of the range of the
luminosity exhibited by magnetic systems  (\citealt{baskill05}; 
\citealt{rana06}; \citealt{byck10}; \citealt{reis13}) and hence may 
make a substantial contribution. Conceivably Galactic scale influences
such as the metallicity or binarity of the stellar population may also
play a role. For example, a significant increase in the number
density of extreme RSCVn binaries towards the inner galaxy (e.g. 
\citealt{rev11}), with spectra dominate by relatively hard ($\sim 3$ keV) 
coronal plasma, could in principle provide the required 
6.7-keV EW enhancement.

\section{Conclusions}
\label{sec_7}

We have shown that the average spectral properties, particularly the Fe-line
properties, of hard X-ray sources discovered serendipitously at intermediate to
faint fluxes in \xmmn observations of the Galactic plane match those of known
magnetic CVs. Unfortunately this X-ray source population is too faint
to be readily identfied via the cross-correlation of the \xmmn positions
with current NIR source catalogues.  However, our analysis does provide an
indication that, in Galactic X-ray surveys extending down
to a modest $ 10^{-13}$  \ergscm (2--10 keV), an appropriate combination of
sub-arcsec X-ray astrometry and deep high-resolution NIR imaging will
reveal substantial numbers of (relatively high-L$_X$) CVs.

Although, the stacked spectra of the hard \xmmn sources show a fair
resemblance to the integrated spectrum of the GRXE, there are some
differences in detail, notably a factor 2-3 discrepancy in the
observed EW of the 6.7-keV He-like Fe line. This discrepancy presumably
stems from the changing makeup of the X-ray source population at a
flux threshold some three orders of magnitude fainter than probed by 
the \xmmn observations.


\section*{Acknowledgements}

This publication makes use of data products from 2MASS, which is a joint
project of the University of
Massachusetts and the Infrared Processing and Analysis
Center/California Institute of Technology, funded by the National
Aeronautics and Space Administration and the National Science
Foundation. In carrying out this research, use has been made of ALADIN, 
VizieR and SIMBAD at the CDS, Strasbourg, France.
The X-ray datasets were obtained from the \xmmn Science Archive
(XSA)\footnote{http://xmm.esac.esa.int/xsa/}.



\appendix
\section{The 138 serendipitous 2XMM sources.}

\begin{table}
\small
\caption{The source sample. The columns provide the following information
(from left to right): the source and observation identification (OBSID) 
numbers, the IAU name, the pn net counts in the source spectrum and the
broad-band hardness ratio (HR).}
\begin{tabular}{ccccc}
\\
\hline
\hline
Source & ObsID & IAU name & pn net counts & HR \\
       &       &  2XMM    &               &    \\ 
\hline					   
210866& 0400910201& J145847.6-581623 & 369 & -0.89\\
211566& 0405390401& J153615.4-575415 & 834 & -0.78\\
142052& 0203910101& J154951.7-541630 & 709 & -0.04\\
142126& 0203910101& J155037.5-540722 & 123 & -0.22\\
212112& 0406650101& J161414.2-514857 & 270 & -0.70\\
212156& 0406650101& J161437.3-512935 & 593 & -0.91\\
212182& 0406650101& J161448.5-514830 & 802 & -0.93\\
212210& 0406650101& J161502.8-513802 & 362 & -0.80\\
212279& 0406650101& J161552.6-513756 & 884 &  0.01\\
212289& 0406650101& J161601.7-513715 & 205 & -0.40\\
212343& 0406750201& J162048.8-494214 & 250 & -0.19\\
212403& 0403280201& J162608.1-490010 & 183 & -0.65\\
148350& 0307170201& J163737.0-472951 & 215 & -0.95\\
148374& 0307170201& J163748.3-472220 & 388 &  0.13\\
148388& 0307170201& J163756.4-471949 & 515 & -0.88\\          
148402& 0307170201& J163802.7-471357 & 239 &  0.98\\          
148411& 0307170201& J163808.5-472607 & 223 &  0.87\\           
148493& 0307170201& J163835.9-472145 & 441 &  0.92\\           
148497& 0303100101& J163837.7-464725 & 118 &  0.94\\           
148502& 0303100101& J163839.2-470618 & 129 &  0.05\\           
148509& 0307170201& J163840.9-471952 & 169 & -0.62\\           
148513& 0307170201& J163842.3-473008 & 257 &  0.95\\           
148559& 0303100101& J163855.1-470146 & 470 & -0.04\\
148620& 0303100101& J163914.5-470020 & 345 & -0.81\\
148624& 0307170201& J163915.1-472310 & 253 & -0.89\\ 
148707& 0307170201& J163937.8-471951 & 118 &  1.00\\
148790& 0303100101& J164004.1-470419 & 283 & -0.88\\
150181& 0112460201& J165340.2-395709 & 139 &  0.40\\
150188& 0112460201& J165341.1-395735 & 122 &  0.19\\
150845& 0112460201& J165443.1-394804 & 172 & -0.20\\
150957& 0112460201& J165515.6-394544 & 224 & -0.60\\
151081& 0200900101& J165739.8-425715 & 437 &  0.95\\
151278& 0200900101& J165906.6-424210 & 275 &  0.07\\
213203& 0406752301& J170019.2-422019 & 392 &  1.00\\
213412& 0406750301& J170451.7-410949 & 178 &  0.41\\
152200& 0144080101& J170713.1-405414 & 208 & -0.70\\
152262& 0144080101& J170819.6-404606 & 124 &  0.94\\
214415& 0401960101& J171808.7-382604 & 375 & -0.82\\
214420& 0401960101& J171813.6-382516 & 355 &  0.16\\
214437& 0401960101& J171830.9-382704 & 225 & -0.79\\
214440& 0401960101& J171833.0-382749 & 246 & -0.92\\
154343& 0112201401& J172945.2-335028 & 152 &  0.26\\
155792& 0112971901& J174351.2-284638 & 178 & -0.58\\
155865& 0112970701& J174423.3-291743 & 569 &  0.03\\ 
215274& 0406580201& J174434.4-301521 & 317 & -0.95\\
215297& 0406580201& J174441.2-301647 & 215 & -0.90\\
156010& 0112970701& J174458.2-292507 & 176 &  0.96\\
156080& 0103261301& J174508.0-303906 & 110 &  0.98\\
215498& 0400340101& J174541.1-300055 & 251 &  0.23\\
215524& 0400340101& J174558.0-295738 & 129 &  0.98\\
157004& 0112970201& J174645.2-281547 & 212 &  1.00\\
157045& 0112970201& J174654.6-281658 & 146 & -0.19\\
157148& 0205240101& J174716.1-281047 & 441 &  0.96\\
157193& 0205240101& J174722.8-280905 & 363 &  0.99\\
157247& 0205240101& J174730.8-281347 & 110 & -0.82\\
157393& 0205240101& J174804.1-281446 & 104 &  0.92\\			
157432& 0205240101& J174814.0-281621 & 62  &  1.00\\
\hline
\hline
\label{table_a1}
\end{tabular}
\end{table}

\begin{table}
\small
\addtocounter{table}{-1}
\centering
\linespread{1}
\begin{tabular}{ccccc}
\\
\hline
\hline
Source & ObsID & IAU name & pn net counts & HR  \\
       &       &  2XMM    &               &     \\
\hline		
157439& 0205240101& J174816.9-280750 & 225 & 0.99\\
157544& 0112970101& J174848.0-281240 & 402&  0.37\\
157571& 0112970101& J174858.3-281422 & 172&  0.23\\
158008& 0206990101& J175035.2-311825 & 197& -0.48\\
158050& 0206990101& J175046.8-311629 & 237& -0.02\\
158316& 0307110101& J175136.4-295016 & 170& -0.77\\
158318& 0307110101& J175137.1-295522 & 300&  0.39\\ 
158373& 0307110101& J175155.9-295113 & 514& -0.38\\
158409& 0307110101& J175233.1-293944 & 203& -0.57\\
158631& 0302570101& J175446.3-285659 & 202&  0.46\\ 
158719& 0206590201& J175516.6-293655 & 243&  0.11\\
158755& 0302570101& J175528.8-290002 & 491&  0.60\\
158765& 0302570101& J175533.4-290148 & 272& -0.81\\
158770& 0302570101& J175534.1-291136 & 394&  0.75\\
158784& 0302570101& J175539.0-285637 & 390& -0.88\\
158786& 0302570101& J175540.7-290641 & 232&  0.12\\
158847& 0302570101& J175607.6-290752 & 220&  0.56\\
158874& 0302570101& J175624.0-290924 & 249& -0.60\\
158926& 0099760201& J175656.5-214833 & 134& -0.70\\
158936& 0099760201& J175659.8-214350 & 806& -0.04\\ 
158945& 0099760201& J175701.8-213336 & 226&  0.03\\
159002& 0099760201& J175722.3-213729 & 231&  0.49\\
159099& 0099760201& J175741.1-214309 & 580& -0.98\\
159225& 0099760201& J175800.6-213856 & 538&  0.37\\
160502& 0135742801& J180154.4-224647 & 258& -0.97\\
162701& 0024940201& J180706.8-192559 & 296& -0.94\\
162842& 0024940201& J180736.4-192658 & 338&  0.06\\
162991& 0024940201& J180802.0-191505 & 325& -0.68\\
163006& 0024940201& J180804.4-192453 & 311& -0.04\\
163107& 0024940201& J180822.4-191813 & 202& -0.65\\
163452& 0301270401& J180929.5-194126 & 310& -0.81\\
163464& 0301270401& J180932.7-194654 & 209& -0.85\\
163510& 0301270401& J180942.9-194544 & 288&  1.00\\
163664& 0301270401& J181009.8-194829 & 357& -0.21\\
163722& 0152835701& J181024.8-183712 & 216&  0.79\\
165111& 0152834901& J181848.0-152753 & 397& -0.91\\	
165128& 0152834501& J181851.2-155920 & 352&  0.85\\
165294& 0152834501& J181913.4-160118 & 320&  0.45\\
165845& 0040140201& J182258.0-135322 & 185&  0.94\\
165849& 0040140201& J182304.3-135036 & 154&  0.90\\
165864& 0040140201& J182319.7-134009 & 275&  0.78\\
165865& 0040140201& J182321.2-134643 & 316& -0.75\\ 
165873& 0040140201& J182342.7-134947 & 510& -0.58\\
165933& 0051940101& J182524.5-114525 & 293&  0.67\\
165958& 0104460701& J182533.6-121452 & 189& -0.89\\ 
166079& 0051940301& J182626.7-112854 & 203&  0.79\\
166111& 0135745401& J182639.6-114216 & 202& -0.75\\
166331& 0104460401& J182756.7-110448 & 174&  0.90\\
166437& 0051940401& J182830.8-114514 & 265& -0.83\\
166459& 0104460401& J182845.5-111710 & 557& -0.81\\
166461& 0135745701& J182847.6-101337 & 551& -0.98\\ 
166477& 0135745701& J182855.9-095414 & 321&  0.88\\ 
166494& 0104460901& J182905.6-104635 & 178&  0.84\\
166583& 0135745801& J182948.6-110604 & 240&  0.46\\
166713& 0135746301& J183038.2-100246 & 223&  0.80\\
166900& 0135741601& J183251.4-100106 & 102&  0.22\\
217348& 0400910101& J183345.2-081828 & 593& -0.95\\
167248& 0302560301& J183514.6-083740 & 168& -0.04\\
217485& 0400910301& J183711.5-063315 & 212& -0.96\\
167514& 0301880901& J183939.8-024935 & 185& -0.27\\ 
167528& 0301880901& J183947.4-024504 & 216& -0.81\\ 
217584& 0302970301& J184226.1-035947 & 403&  0.99\\
\hline
\hline
\end{tabular}
\end{table}

\begin{table}
\small
\addtocounter{table}{-1}
\centering
\linespread{1}
\begin{tabular}{ccccc}
\\
\hline
\hline
Source & ObsID & IAU name & pn net counts & HR \\
       &       &  2XMM    &           &        \\
\hline		
217586& 0302970301& J184226.9-035536 & 181 & -0.47\\
168031& 0046540201& J184441.9-030551 & 466 &  0.48\\
168039& 0017740601& J184447.7+011131 & 316 & -0.86\\
168124& 0203850101& J184720.6-015248 & 227 &  0.98\\
168158& 0207010201& J184747.3-011910 & 149 &  0.95\\
217784& 0302970801& J184805.8-022821 & 231 & -0.03\\ 
217813& 0302970801& J184816.8-022524 & 436 &  0.97\\
217816& 0406140201& J184817.4-031907 & 675 &  0.89\\
168249& 0136030101& J184953.6-003007 & 355 & -0.60\\
168254& 0136030101& J184958.7-002018 & 170 &  0.58\\
168272& 0136030101& J185020.0-001313 & 449 & -0.83\\
168311& 0017740401& J185125.1+000742 & 182 & -0.94\\
168323& 0017740401& J185139.1+001635 & 378 & -0.09\\
168324& 0017740401& J185139.9+001308 & 191 & -0.95\\
168428& 0017740401& J185233.2+000638 & 189 & -0.14\\
169142& 0136030201& J190109.0+045751 & 303 &  0.61\\
169450& 0305580201& J190704.0+092532 & 172 &  0.56\\ 
169482& 0305580101& J190717.7+092421 & 288 &  0.42\\
169532& 0305580101& J190742.0+090713 & 270 &  0.83\\
\hline
\hline
\end{tabular}
\end{table}

\label{lastpage}


\begin{thebibliography}{99}
\bibitem[\protect\citeauthoryear{Agrawal, Riegler, \& White}{1981}]{agr81}
Agrawal P.~C., Riegler G.~R., White N.~E., 1981, MNRAS, 196, 73

\bibitem[\protect\citeauthoryear{Ak et al.}{2007}]{ak07} Ak, T., Bilir, S., Ak, S., Retter, A., 2007, New Ast., 12, 446

\bibitem[\protect\citeauthoryear{Ak et al.}{2008}]{ak08} Ak, T., Bilir, S., Ak, S., Eker Z., 2008, New Ast., 13, 133

\bibitem[\protect\citeauthoryear{Anders \& Grevesse}{1989}]{anders89} Anders E., Grevesse N., 1989, Geochimica et Cosmochimica Acta, 53, 197

\bibitem[\protect\citeauthoryear{Anderson et al.}{2011}]{ander11}
Anderson G.E. et al., 2011, ApJ, 727, 105 
 
\bibitem[\protect\citeauthoryear{Arnaud}{1996}]{arn96} 
Arnaud K., 1996, in Jacoby G. H., Barnes J., eds, ASP Conf. Ser. Vol. 101,
Astronomical Data Analysis Software and Systems V. Astron. Soc. Pac.,
San Francisco, p. 17

\bibitem[\protect\citeauthoryear{Baskill et al.}{2005}]{baskill05}
Baskill D.~S., Wheatley P.~J., Osborne J.~P., 2005, MNRAS, 357, 626

\bibitem[\protect\citeauthoryear{Bernardini et al.}{2012}]{bern12}
Bernardini F., de Martino D., Falanga M., Mukai K., Matt G., 
Bonnet-Bidaud J.-M., Masetti N., Mouchet M., 2012, A\&A, 542, A22

\bibitem[\protect\citeauthoryear{Byckling et al.}{2010}]{byck10}
Byckling K., Mukai K., Thorstensen J.~R., Osborne J.~P., 2010, MNRAS, 408, 2298

\bibitem[\protect\citeauthoryear{Capelli et al.}{2012}]{capelli12}
Capelli, R., Warwick, R. S., Porquet, D., Gillessen, S., Predehl, P.,
2012, A\&A, 545, A35

\bibitem[\protect\citeauthoryear{Chaudhary et al.}{2012}]{chaud12}
Chaudhary, P., Brusa M., Hasinger G., Merloni A., Comastri A., Nandra K.,
2012, A\&A, 537, A6

\bibitem[\protect\citeauthoryear{Covey et al.}{2007}]{covey07} Covey, K.~R. et al., 2007, AJ, 134, 2398

\bibitem[\protect\citeauthoryear{Cropper et al.}{1998}]{cropper98}
Cropper M., Ramsay G.,  Wu K., 1998, MNRAS, 293, 222

\bibitem[\protect\citeauthoryear{Cropper et al.}{1999}]{cropper99}
Cropper M., Wu K., Ramsay G., Kocabiyik A., 1999, MNRAS, 306, 684

\bibitem[\protect\citeauthoryear{Cutri et al.}{2003}]{cutri03} Cutri, R.~M., et al., 2003, Explanatory Supplement to the 2MASS All Sky Data Release and Extended Mission Products (Pasadena:IPAC/Caltech) 

\bibitem[\protect\citeauthoryear{Ebisawa et al.}{2003}]{ebi03}
Ebisawa K., Bourban G., Bodaghee A., Mowlave N., Courvoisier T.~J.~-L.,
2003, A\&A, 411, 59-62

\bibitem[\protect\citeauthoryear{Ebisawa et al.}{2005}]{ebi05}
Ebisawa K. et al., 2005, ApJ, 635, 214

\bibitem[\protect\citeauthoryear{Ebisawa et al.}{2008}]{ebi08}
Ebisawa K. et al., 2008, PASJ, 60, S223

\bibitem[\protect\citeauthoryear{Ezuka \& Ishida}{1999}]{ezuka99}
Ezuka H., Ishida M., 1999, ApJS, 120, 277

\bibitem[\protect\citeauthoryear{Fukazawa}{2011}]{fuk11}
Fukazawa Y. et al., 2011, ApJ, 727, 19

\bibitem[\protect\citeauthoryear{Grimm, Gilfanov \& Sunyaev}{2002}]{grimm02}
Grimm, H.~-J., Gilfanov, M., Sunyaev, R., 2002, A\&A, 391, 923

\bibitem[\protect\citeauthoryear{G{\"u}del}{2004}]{gudel04} G{\"u}del, M., 2004, Astron Astrophys Rev, 12, 71

\bibitem[\protect\citeauthoryear{Hands et al.}{2004}]{hands04}
Hands A.~D.~P., Warwick R.~S., Watson M.~G., Helfand D.~J., 2004, MNRAS, 351, 31

\bibitem[\protect\citeauthoryear{Heard \& Warwick}{2013}]{heard13}
Heard V., Warwick R.~S., 2013, MNRAS, 428, 3462

\bibitem[\protect\citeauthoryear{Hellier, Mukai \& Osborne}{1998}]{hel98}
Hellier C., Mukai K., Osborne J.~P., 1998, MNRAS, 297, 526

\bibitem[\protect\citeauthoryear{Hellier \& Mukai}{2004}]{hel04}
Hellier C., Mukai K., 2004, MNRAS, 352, 1037

\bibitem[\protect\citeauthoryear{Hertz \& Grindlay}{1984}]{hertz84}
Hertz P., Grindlay J.~E., 1984, ApJ, 278, 137

\bibitem[\protect\citeauthoryear{Hong et al.}{2004}]{hong04}
Hong J., Schegel, E.M., Grindlay J.~E., 2004, ApJ, 614, 508

\bibitem[\protect\citeauthoryear{Hong et al.}{2009}]{hong09}
Hong J., van den Berg M., Grindlay J.~E., Laycock S., 2009, ApJ, 706, 223

\bibitem[\protect\citeauthoryear{Hong}{2012}]{hong12}
Hong J., 2012, MNRAS, 427, 1633

\bibitem[\protect\citeauthoryear{Kaneda et al.}{1997}]{kan97}
Kaneda H., Makishima K., Yamauchi S., Koyama K., Matsuzaki K., Yamasaki N.~Y., 1997, ApJ, 491, 638

\bibitem[\protect\citeauthoryear{Koyama et al.}{1986a}]{koy86a}
Koyama K., Makishima K., Tanaka Y., Tsunemi H., 1986a, PASJ, 38, 121

\bibitem[\protect\citeauthoryear{Koyama et al.}{1986b}]{koy86b}
Koyama K., Ikeuchi S., Tomisaka K., 1986b, PASJ, 38, 503

\bibitem[\protect\citeauthoryear{Koyama et al.}{1996}]{koy96}
Koyama K., Maeda, Y., Sonobe T., Takeshima T., Tanaka Y., Yamauchi, S., 
1996, PASJ, 48, 249

\bibitem[\protect\citeauthoryear{Koyama et al.}{2007}]{koy07}
Koyama K. et al., 2007, PASJ, 59, 245

\bibitem[\protect\citeauthoryear{Krivonos et al.}{2007}]{kriv07}
Krivonos R., Revnivtsev M., Churazov E., Sazonov S., Grebenev S.,
Sunyaev R., 2007, A\&A, 463, 957

\bibitem[\protect\citeauthoryear{Lawrence et al.}{2007}]{law07} Lawrence A.
et al., 2007, MNRAS, 379, 1599

\bibitem[\protect\citeauthoryear{Laycock et al.}{2005}]{laycock05}
Laycock, S., Grindlay, J., van den Berg, M., Zhao, P., Hong, J., Koenig, X., 
Schlegel, E.M., Persson, S.E., 2005, ApJ, 634, L53

\bibitem[\protect\citeauthoryear{Liu et al.}{2007}]{liu07}
Liu Q.Z., van Paradijs J., van den Heuvel E.P.J., 2007, A\&A, 469, 807

\bibitem[\protect\citeauthoryear{Mauerhan et al.}{2009}]{mau09}
Mauerhan J.~C., Muno M.~P., Morris M.~R., Bauer F.~E., Nishiyama S., Nagata T., 2009, ApJ, 703,30 

\bibitem[\protect\citeauthoryear{Mauerhan et al.}{2010}]{mau10}
Mauerhan J.~C., Muno M.~P., Morris M.~R., Stolovy S.~R., Cotera A., 2010, ApJ, 710, 706

\bibitem[\protect\citeauthoryear{Morihana et al.}{2013}]{morihana13}
Morihana K., Tsujimoto, M., Yoshida T., Ebisawa K., 2013, ApJ, 766, 14

\bibitem[\protect\citeauthoryear{Motch et al.}{2010}]{motch10} Motch, C.
et al., 2010, A\&A, 523, A92

\bibitem[\protect\citeauthoryear{Muno et al.}{2004}]{muno04}
Muno, M.P. et al., 2004, ApJ, 613, 1179

\bibitem[\protect\citeauthoryear{Nebot G\'{o}mez-Mor\'{a}n et al.}{2013}]{nebot13}
Nebot G\'{o}mez-Mor\'{a}n, A. et al., 2013, A\&A, 553A, 12N

\bibitem[\protect\citeauthoryear{Nobukawa et al.}{2009}]{nob09}
Nobukawa, M., Koyama, K., Matsumoto, H., Tsuru, T.G., 2009, PASJ, 61, 93  

\bibitem[\protect\citeauthoryear{Nishiyama et al.}{2013}]{nish13}
Nishiyama S. et al., 2013, ApJL, 769, L28

\bibitem[\protect\citeauthoryear{Rana et al.}{2006}]{rana06}
Rana V.R., Singh K.~P., Schegel E.~M., Barrett P.~E., 2006, ApJ, 1042

\bibitem[\protect\citeauthoryear{Reis et al.}{2013}]{reis13}
Reis R.~C., Wheatley P.~J., G\"{a}nsicke, B.~T., Osborne J.~P., 2013, MNRAS, 430, 1994 

\bibitem[\protect\citeauthoryear{Revnivtsev et al.}{2006}]{rev06}
Revnivtsev M., Sazonov S., Gilfanov M., Churazov E., Sunyaev R., 2006, A\&A, 452, 169

\bibitem[\protect\citeauthoryear{Revnivtsev et al.}{2008}]{rev08}
Revnivtsev M., Sazonov S., Krivonos R., Ritter H., Sunyaev R., 2008, A\&A, 489, 1121

\bibitem[\protect\citeauthoryear{Revnivtsev et al.}{2009}]{rev09}
Revnivtsev M., Sazonov S., Churazov E., Forman W., Vikhlinin A., Sunyaev R., 2009, Nature, 458, 1142

\bibitem[\protect\citeauthoryear{Revnivtsev et al.}{2011}]{rev11}
Revnivtsev M., Sazonov S., Forman W., Churazov E., Sunyaev R., 2011, MNRAS, 414, 495

\bibitem[\protect\citeauthoryear{Ritter \& Kolb}{2003}]{ritter03}
Ritter H., Kolb, U., 2003, A\&, 404, 301

\bibitem[\protect\citeauthoryear{Ruiter et al.}{2006}]{ruiter06}
Ruiter A.J., Belczynski K., Harrison T. E., 2006, ApJ, 640, L167

\bibitem[\protect\citeauthoryear{Sazonov et al.}{2006}]{saz06}
Sazonov S., Revnivtsev M., Gilfanov M., Churazov E., Sunyaev R., 2006, A\&A, 450, 117

\bibitem[\protect\citeauthoryear{Sidoli et al.}{2006}]{sidoli06}
Sidoli, L., Merghetti, S., Favata, F., Oosterbroek, T.,  
Parmar, A.N., 2006, A\&A 456, 287
                     
\bibitem[\protect\citeauthoryear{Skrutskie et al.}{2006}]{skrutskie06} Skrutskie, M.~F. et al., 2006, AJ, 131, 1163

\bibitem[\protect\citeauthoryear{Strickland et al.}{2000}]{str00}
Strickland D.~K., Heckman T.~M., Weaver K.~A., Dahlem, M.,
2000, AJ, 120, 2965

\bibitem[\protect\citeauthoryear{Strong et al.}{2005}]{strong05}
Strong A.W., Diehl R., Halloin H., Schönfelder V., Bouchet L.,
Mandrou P., Lebrun F., Terrier R., 2005, A\&A 444, 495

\bibitem[\protect\citeauthoryear{Sugizaki et al.}{2001}]{sug01}
Sugizaki M., Mitsuda K., Kaneda H., Matsuzaki K., Yamauchi S., Koyama K., 2001, ApJS, 134, 77

\bibitem[\protect\citeauthoryear{Tanaka}{2002}]{tanaka02}
Tanaka, Y., 2002, A\&A, 382, 1052

\bibitem[\protect\citeauthoryear{Tanaka \& Yamauchi}{2010}]{tanaka10}
Tanaka, Y., Yamauchi, S., 2010, in  Makishima K., ed.,
The Energetic Cosmos: from Suzaku to ASTRO-H,  
Proceedings of the 3rd Suzaku Conference, 2009, 
Otaru, Japan. JAXA Special Publication JAXA-SP-09-008E. 

\bibitem[\protect\citeauthoryear{Uchiyama et al.}{2011}]{uch11}
Uchiyama, H., Nobukawa, M., Tsuru, T. G., Koyama, K., Matsumoto, H., 2011,
PASJ, 63, 903

\bibitem[\protect\citeauthoryear{Uchiyama et al.}{2013}]{uch13}
Uchiyama, H., Nobukawa, M., Tsuru, T. G., Koyama, K., 2013,
PASJ, 65, 19

\bibitem[\protect\citeauthoryear{Valinia \& Marshall}{1998}]{val98}
Valinia A., Marshall F.~E., 1998, ApJ, 505, 134

\bibitem[\protect\citeauthoryear{Valinia et al.}{2000}]{val00}
Valinia A., Tatischeff V., Arnaud K., Ebisawa K., Ramaty R., 2000, ApJ, 543, 733

\bibitem[\protect\citeauthoryear{van den Berg et al.}{2009}]{berg09}
van den Berg M., Hong J., Grindlay, J.~E., 2009, ApJ, 700, 1702

\bibitem[\protect\citeauthoryear{van den Berg et al.}{2012}]{berg12}
van den Berg M., Penner Kyle, Hong J.,  Grindlay J.~E., Zhao P.,
Laycock S.,  Servillat M., 2012, ApJ, 748, 31
 
\bibitem[\protect\citeauthoryear{Warwick et al.}{1985}]{war85}
Warwick R.~S., Turner M.~J.~L., Watson M.~G., Willingale R., 1985, Nature, 317, 218

\bibitem[\protect\citeauthoryear{Warwick et al.}{2011}]{war11}
Warwick R.~S., P\'erez-Ram\'irez D., Byckling K., 2011, MNRAS, 413, 595

\bibitem[\protect\citeauthoryear{Watson et al.}{2009}]{wat09}
Watson M.~G. et al., 2009, A\&A, 493, 339

\bibitem[\protect\citeauthoryear{Worrall et al.}{1982}]{wor82}
Worrall D.~M., Marshall F.~E., Boldt E.~A., Swank J.~H., 1982, ApJ, 255, 111

\bibitem[\protect\citeauthoryear{Yamasaki et al.}{1997}]{yam97}
Yamasaki N.~Y. et al., 1997, ApJ, 481, 821

\bibitem[\protect\citeauthoryear{Yamauchi \& Koyama}{1993}]{yam93}
Yamauchi S., Koyama K., 1993, ApJ, 404, 620

\bibitem[\protect\citeauthoryear{Yamauchi et al.}{1996}]{yam96}
Yamauchi, S.,  Kaneda, H., Koyama, K., Makishima, K., Matsuzaki, K.,
Sonobe, T., Tanaka, Y., Yamasaki, N., 1996, PASJ, 48, L15

\bibitem[\protect\citeauthoryear{Yamauchi et al.}{2009}]{yam09}
Yamauchi S., Ebisawa K., Tanaka Y., Koyama K., Matsumoto H., 
Yamasaki N.~Y., Takahashi H., Ezoe Y., 2009, PASJ, 61, S225

\bibitem[\protect\citeauthoryear{Yuasa et al.}{2010}]{yuasa10}
Yuasa T., Nakazawa K., Makishima K., Saitou K., Ishida M., Ebisawa K.,
Mori H., Yamada S., 2010, A\&A, 520, A25

\bibitem[\protect\citeauthoryear{Yuasa et al.}{2012}]{yuasa12}
Yuasa T., Makishima K., Nakazawa K.,  2012, ApJ, 753, 129


\end{thebibliography}
\end{document}